\documentclass[a4paper,fleqn,usenatbib]{mnras}

\usepackage[T1]{fontenc}
\usepackage{ae,aecompl}
\usepackage{appendix}
\usepackage{graphicx}	
\usepackage{amsmath}	
\usepackage{amssymb}	
\usepackage{booktabs}	
\usepackage{csquotes}	
\usepackage[normalem]{ulem}
\usepackage{url}
\usepackage{caption}

\usepackage{newtxtext,newtxmath}

\usepackage{tablefootnote,footnotehyper} 
\usepackage{array}
\newcolumntype{H}{>{\setbox0=\hbox\bgroup}c<{\egroup}@{}} 

\title[Population synthesis for A-type stars]{Quantifying isochrone‑based age uncertainties for rapidly rotating A‑type stars}

\author[Murphy et al.]{
Simon J. Murphy,$^{1}$\thanks{E-mail: simon.murphy@unisq.edu.au}
Anuj Gautam,$^{1}$\thanks{E-mail: anuj.gautam@unisq.edu.au}
and Zachary R. Claytor,$^{2}$\thanks{E-mail: zclaytor@stsci.edu}
\\
$^{1}$ Centre for Astrophysics, University of Southern Queensland, Toowoomba, QLD 4350, Australia\\
$^{2}$ Space Telescope Science Institute, 3700 San Martin Drive, Baltimore, MD 21218, USA\\
}

\date{Accepted XXX. Received YYY; in original form ZZZ}
\pubyear{2025}

\begin{document}
\label{firstpage}
\pagerange{\pageref{firstpage}--\pageref{lastpage}}
\maketitle

\begin{abstract}
Accurate stellar ages and masses are essential for interpreting the demographics and physical properties of exoplanets, particularly for intermediate-mass, early-type stars where conventional age indicators are ineffective. Isochrone fitting remains the primary tool for characterising such stars, yet its uncertainties are often underestimated, especially in the presence of rapid rotation and unresolved binarity. 
We present a population‑synthesis framework designed to quantify realistic mass and age uncertainties for intermediate‑mass stars (1.4--2.5\,M$_{\odot}$), incorporating distributions in rotation rate, mass, metallicity, binarity, inclination, and observational error. Rotational and geometric effects are applied {\it a posteriori} to stellar evolutionary models, enabling a continuous treatment of rotation and its impact on effective temperature and luminosity.
By comparing synthetic populations against commonly used isochrone grids, we demonstrate that rotation and unresolved companions systematically bias inferred masses and ages, particularly for young stars, and introduce random uncertainties at the $\sim$0.1-M$_{\odot}$ and $\sim$180-Myr level, often exceeding formal fitting errors. The effect is strongest near the zero-age main sequence, where ages are underestimated by a factor of $\geq2$, while for older A stars ($>$10\% of their main-sequence lifetime), ages are overestimated by 31\% with 27\% scatter. Our findings carry important consequences for planet detectability, characterisation, and population studies.
We provide a publicly available tool, RAPID, for probabilistic inference of stellar parameters from these synthetic populations, and we demonstrate its application to known exoplanet hosts.
\end{abstract}

\begin{keywords}
binaries: general -- stars: fundamental parameters -- stars: early-type -- stars: rotation -- stars: statistics
\end{keywords}



\section{Introduction}
\label{sec:intro}

Isochrones serve as a cornerstone for characterizing the masses and ages of stars and their planets. Those stellar masses and ages are also important predictors of planet detection outcomes \citep{johnsonetal2007b,swastiketal2024}, especially for direct imaging \citep{nielsenetal2019b}. This is because planets are at their most luminous at young ages \citep{bowler2016}, and the giant planets that are amenable to direct imaging are intrinsically more common around more massive ($M>1.5$\,M$_{\odot}$) stars \citep{lannieretal2016}.
Yet the yields from direct-imaging surveys optimised by host star mass and age are still relatively low at $<$10\% \citep{nielsenetal2019b}, so it is important that the target stars' ages are correct.

While several studies have evaluated the reliability of ages of exoplanet hosts \citep{bonfantietal2015,bonfantietal2016,silvaaguirreetal2015,christensen-dalsgaard&aguirre2018,bergeretal2023,swastiketal2023}, including the impact of using different age estimators, most have focused on FGK-type stars; few have focused on early-type stars \citep{david&hillenbrand2015}, where age accuracy is most crucial. 

Both the mass and age of a directly imaged exoplanet are affected by uncertainty in the age of the host star. Only the planet's luminosity, not its mass, is measured by direct imaging, hence, one must resort to luminosity evolution models to infer the planet's mass \citep{currieetal2023a}. These models relate the luminosity of a planet to its mass at the estimated stellar age. Hence, the importance of age uncertainty transcends planet detectability to planet characterization and demographics \citep{barberetal2024,squicciarinietal2025}.

The pace of exoplanet evolution is particularly rapid in the first 50\,Myr. The combined effects of atmospheric boil-off, Kelvin-Helmholz contraction, and XUV-driven photoevaporation can shrink the radii of Neptune-mass planets from $\sim$10\,R$_{\rm E}$ to 4\,R$_{\rm E}$ in this time-frame \citep[e.g.][]{owen&wu2013,lopez&fortney2013,rogers&owen2021,rogersetal2024a}. Interpretation of the physical properties for such young planets, as well as their atmospheric compositions \citep{gupta&schlicting2020}, therefore depend on accurate ages. This is especially true for discoveries in complex star-forming regions, such as the Sco-Cen association \citep[e.g.][]{ratzenbocketal2023}, for which a significant transiting planet population has been found thanks to the NASA K2 and TESS missions \citep{mannetal2016,davidetal2016,rizzutoetal2020,mannetal2022,zakhozhayetal2022,vachetal2025}.

Accurate stellar ages are of astrophysical importance beyond their application to young exoplanet hosts. In Galactic archaeology, accurate ages constrain the trace-back of stellar orbits in the Milky Way to understand its stellar populations \citep{nessetal2019,kordopatisetal2023}. Similarly, in planetary system dynamics, reliable stellar ages anchor models of long-term orbital evolution. Processes such as obliquity excitation (including via binary companions; \citealt{su&dong2025}), tidal dissipation, and secular interactions operate over gigayear timescales \citep{albrechtetal2012,knudstrupetal2024,duganetal2025}, but our understanding of these processes is limited by our ability to assign ages to exoplanet systems of a variety of ages \citep{albrechtetal2022}. 

Star clusters and stellar associations offer some of the best-determined ages for any type of star, because one can model the age of the whole population rather than fitting individual stars (e.g. \citealt{grattonetal2024}). However, the existence of age gradients across even the youngest of star-forming regions \citep{pecaut&mamajek2016}, intrinsic age spreads of up to 5\,Myr \citep{soderblom2010,krumholtzetal2019}, and variable pre-main-sequence lifetimes due to bursty accretion \citep{vorobyov&elbakyan2020,dasetal2025}, all make association ages less dependable. Asteroseismic ages are the least affected by observational uncertainty and also happen to have been obtained for direct imaging targets \citep{murphyetal2021a,currieetal2023a}, but such analyses are time-consuming and not all stars have suitable pulsation spectra. Hence, isochrones remain the de facto tool for stellar age determination, especially for early-type stars where gyrochronology and lithium depletion are not applicable \citep{soderblom2010}.


Isochrones are constructed by interpolating stellar evolutionary tracks, hence they are highly sensitive to the choice of model physics \citep{lebretonetal2014a}. They allow masses and ages to be inferred from stellar effective temperatures and luminosities, which are now available for $>$$10^8$ Milky Way stars \citep{fouesneauetal2023,mcdonaldetal2025}. \citet{tayaretal2022}   highlighted that systematic uncertainties affecting those observables, such as the calibration uncertainty of effective temperature via stellar interferometry (see also \citealt{casagrandeetal2014}), limit the achievable isochrone age accuracy to about 20\%, with the actual uncertainties often being larger. Their study primarily focused on FGK stars.


In this work, we focus on realistic age and mass uncertainties for predominantly A-type, intermediate-mass stars (1.4--2.5\,M$_{\odot}$). Primarily due to their rapid rotation, such stars are understudied despite 
embodying the stellar mass range with the highest occurrence rate of giant planets \citep{johnsonetal2010,reffertetal2015}. Our approach is agnostic to the presence of planets and not limited to young stars; hence, the results will be useful to characterization of the hosts of planets discovered by any means, and indeed to stellar astrophysics more broadly. We pay particular attention to the effects of rotation, unresolved binarity, and random measurement uncertainty.

Our approach is to use stellar population synthesis \citep[e.g.][]{toonenetal2014,zonoozietal2025}. 
Population synthesis models have previously been constructed for colour-magnitude diagrams to quantify the impact of rotationally-induced gravity and limb darkening on the appearance and inferred ages of star clusters \citep{georgyetal2014}. That work framed subsequent discussion around rotation as the cause of extended main-sequence turn-offs (e.g. \citealt{niederhoferetal2015,brandt&huang2015c,bastianetal2016,goudfrooijetal2017,limetal2019,sunetal2019}). Here, we create population synthesis models with the addition of random measurement uncertainty to quantify isochrone-fitting uncertainties in a different context: the application to field stars, especially exoplanet host stars. We use realistic input distributions across mass, metallicity, age, rotation, and binarity, and provide public tools for inferring these properties based on effective temperature and luminosity (and optionally other parameters).

We describe the population synthesis and underlying physical parameter distributions in Sec.\,\ref{sec:synth}, with particular focus on the treatment of rotation.
Sec.\,\ref{sec:applications} is dedicated to applications. We compare our synthetic population against different sets of public isochrones, including those constructed from different stellar rotation rates.
We show how masses and ages can be inferred from our synthetic population using HD\,250208 (TOI\,2497) and HD\,56414 as examples, then we use KELT-20 (MASCARA-2) to show how full posterior distributions of stellar parameters can be obtained using RAPID, an applet that we make available for public use. Our conclusions are given in Sec.\,\ref{sec:conclusions}.

\section{Population synthesis}
\label{sec:synth}

We have synthesised a population of intermediate-mass (1.4--2.5\,M$_{\odot}$) stars by taking existing stellar models and modifying them to follow the desired properties at a population level. The idea is to have realistic distributions in mass, metallicity, age, and rotation, also allowing for stars to have binary companions. Our lower mass bound ensures that our stars lie above the Kraft break (which is at 1.32--1.41\,M$_{\odot}$; \citealt{beyer&white2024}), and are thus rapid rotators. Our approach is to calculate non-rotating models then apply the effects of rotation and companions {\it a posteriori}.

\begin{figure*}
\begin{center}
\includegraphics[width=0.65\textwidth]{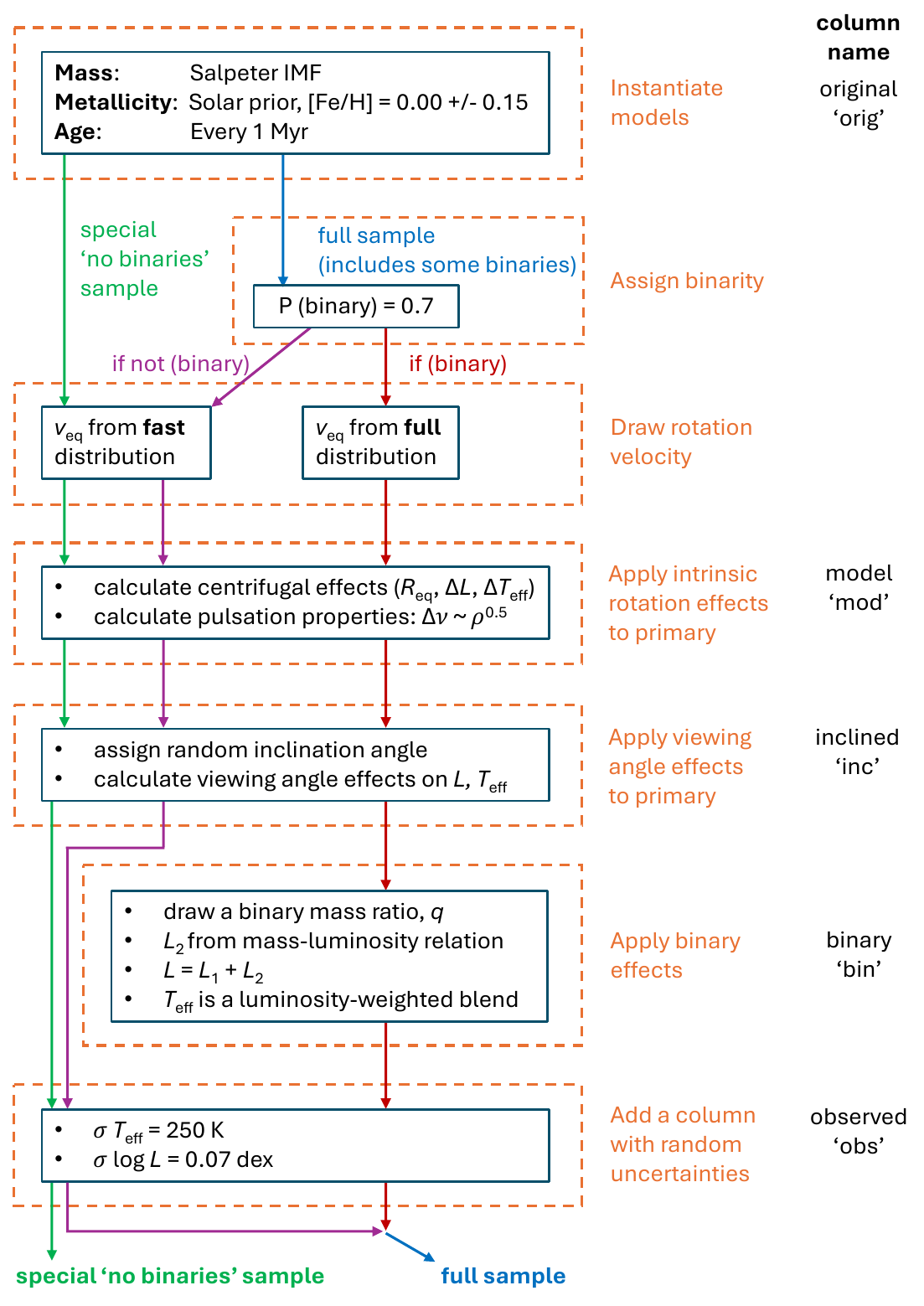}
\caption{Workflow diagram summarising the various steps in the population synthesis, and their relation to the outputted $T_{\rm eff}$ and $L$ columns (whose names are given in black, right-hand side) in both of the available output files: {\tt no\_binaries} and {\tt full}. Further details are given in Sec.\,\ref{ssec:summary}.}
\label{fig:overview}
\end{center}
\end{figure*}

The underlying stellar evolutionary and pulsation models have been calculated with {\sc mesa} and {\sc gyre} by \citet{gautametal2026}, where detailed descriptions of the model physics are given. Here, we focus only on population-level effects. The order of some of the operations below is important (particularly Sec.\,\ref{ssec:rotation_and_binarity} and \ref{ssec:obs_uncertainty}), should readers wish to synthesise similar populations. A graphical overview is provided in Fig.\,\ref{fig:overview} and a summary is given at the end of this section (Sec.\,\ref{ssec:summary}).

\subsection{Initial sample selection}
\label{ssec:sample}

We drew our initial sample from the models of \citet{gautametal2026}. We used all main-sequence (MS) models with zero rotational velocity, because we implemented our own prescription of rotation (Sec.\,\ref{sssec:rotation_draw}) where we use a continuous distribution of rotation rates rather than the discrete values in \citeauthor{gautametal2026}. The {\it initial} model selection spans a mass range of $M$ = 1.4 to 2.5 in steps of 0.02\,M$_{\odot}$ and metal mass fractions $Z$ = 0.001 to 0.026 in steps of 0.001, which corresponds to [Fe/H] = $-1.16$ to $+0.25$. This parameter space is similar to the non-rotating models of \citet{murphyetal2023}, which would also have served our purposes. However, we chose the \citet{gautametal2026} models to benchmark our rotation prescription. All models had core overshooting of {\tt overshoot\_f} $= 0.017$ and {\tt overshoot\_f0} $= 0.002$ in the {\sc mesa} prescription \citep{jermynetal2023}.

After this initial selection, we sample the grid in various ways based on age, mass, and metallicity, before applying other effects.

\subsection{Age selection and sampling}
\label{ssec:age}
The evolutionary tracks in \citet{gautametal2026} are sampled more densely (shorter time-steps) on the pre-MS up to young MS ages of 50\,Myr to maintain accuracy, then the sampling density is lowered for the MS where stellar properties change slowly. If these models were sampled directly for population synthesis, pre-MS stars would be overrepresented in our sample. 
To solve this and ensure homogeneous sampling, we interpolated along each evolutionary track to sample stellar properties at 1-Myr intervals. The resulting models span evolutionary stages from the pre-MS phase through to the post-MS contraction phase, which is sometimes called the `hook' of the terminal age main sequence (TAMS).
The latter phase is very short, representing only $\sim$1.6\% of the main-sequence lifetime of a solar-metallicity model at 1.7\,M$_{\odot}$, and $\sim$0.6\% at 2.4\,M$_{\odot}$.\footnote{This suggests that the large number of papers in the literature claiming to observe stars in the post-MS phase might have failed to account for this low prior probability. These time-frames depend on core overshooting properties, which were not varied in this study.}

The 1-Myr sampling was initially applied regardless of stellar mass, which means that stars that live longer are represented by a greater number of models in the population. This gives a realistic impression of stellar populations. To give an asteroseismic example, a 1.7-M$_{\odot}$ star moves more slowly through the $\delta$\,Scuti instability strip than a 2.0-M$_{\odot}$ star, hence a snapshot of the instability strip should contain a greater number of 1.7 than 2.0-M$_{\odot}$ stars, regardless of the relative number of lower mass stars in the Galaxy (i.e. regardless of the stellar initial mass function, IMF, which we handle separately).


\subsection{Sampling in mass}
\label{ssec:mass}
Although the mass range of the evolutionary tracks spanned from 1.4 to 2.5\,M$_{\odot}$, it did so uniformly, in steps of 0.02\,M$_{\odot}$. 
To correct this to reflect the IMF, we downsampled the more massive stars in the grid. At masses above 1.0\,M$_{\odot}$, the IMF scales as $(m$/M$_{\odot})^{-2.35}$ \citep{salpeter1955,kroupa2002,chabrier2003}. Thinning was performed probabilistically across the whole grid according to mass (and later metallicity, see Sec.\,\ref{ssec:metallicity}), rather than thinning every $n^{\rm th}$ model along evolutionary tracks. In this way, each model's inclusion is independent of others on the same evolutionary track. To be specific, for each model we evaluate the quantity $P(m) = (m/1.4)^{-2.35}$, where 1.4 is the lower mass range of the grid, and we keep this model if $P(m)>r_{\rm m}$, where $r_{\rm m}$ is a uniform random number between 0 and 1.

\subsection{Sampling in metallicity}
\label{ssec:metallicity}

Rather than use the full metallicity range of the model grid, we wanted to emulate the distribution of `nearby' field star metallicities. The first step for this was to determine the right prior.

\subsubsection{Determining a metallicity prior}
We first queried the Gaia DR3 source catalogue for stars within $\sim$1\,kpc (specifically, parallax $>$ 1\,mas), so as to avoid extremely metal-poor stars of the halo and bulge, and to avoid metallicity gradients along the Galactic plane with radius \citep[e.g.][]{casagrandeetal2011}. We also limited our search to stars brighter than $G=10$\,mag, so as to capture the types of stars observed by TESS. We did not restrict our results by spectral type, because there are very few A stars with Gaia RVS spectra. Unfortunately, the stars with available spectra are biased towards late spectral types, which naturally includes a somewhat older population than the population of A stars we are interested in (low-mass stars live longer), and a correspondingly sub-solar metallicity distribution (they formed long ago when the galaxy was less metal-rich). Such biases against metal-rich stars in stellar samples in the solar-neighbourhood are well documented \citep[e.g.][]{haywood2001}. Specifically, the above selection criteria yielded a median [M/H] = $-0.19$\,dex with a standard deviation of 0.24\,dex. Hence, although this is in broad agreement with the metallicity distribution of Milky Way dwarfs found in other surveys \citep[e.g.][]{huangetal2022}, those surveys have the same biases, namely that older redder stars outnumber younger bluer stars. For comparison, young, nearby open clusters have metallicities around solar (the Pleiades has [Fe/H] = $0.03\pm0.05$; \citealt{soderblometal2009}). Clearly an alternative metallicity prior that is suitable for A stars is required.

Asteroseismology of young red giants (age < 1\,Gyr) at similar galactocentric radii to the Sun can also yield metallicities that should be similar to the intermediate-mass stars in our simulation. \citet{willettetal2023} constructed a hierarchical Bayesian model of such a sample to constrain the metallicity gradient in the Milky Way thin disc. Galactic chemical enrichment over the past Gyr in the vicinity of the Sun is negligible (e.g. \citealt{daltioetal2021}), and choosing a young sample also limits the resulting influence of radial migration through the Milky Way disc, hence the \citet{willettetal2023} results should be appropriate to this work. The relevant metallicity distribution from their table~1 is [Fe/H] $= -0.022 \pm 0.115$. We use these values as our prior to downselect models in this work, although we note that the uncertainty is smaller than other studies that covered the same parameter space (e.g. \citealt{buderetal2019}, $\sigma=0.17$). Most surveys with results that avoid old metal-poor stars indicate that a Gaussian prior centred on [Fe/H]$\sim$0 is broadly appropriate.

\subsubsection{Implementing the prior}
The prior was implemented by calculating the quantity
$P(Z) = {\rm exp}[-n^2/2]$,
where $n = [\log(Z/Z_{\odot})-\mu_{\rm FeH}] / \sigma_{\rm FeH}$,
and keeping models where $r_{\rm Z} < P(Z)$, where $r_{\rm Z}$ is a uniform random number between 0 and 1 drawn independently from $r_{\rm m}$. For reference, the \citet{gautametal2026} grid of models that we use adopted the \citet{asplundetal2009} solar abundances, wherein the metal mass fraction of the Sun is $Z_{\odot} = 0.0142$.

Our underlying evolutionary tracks were sampled evenly in $Z$ rather than in $\log Z$, hence the density of tracks is greater at $Z>Z_{\odot}$ than at $Z<Z_{\odot}$. It was therefore necessary to confirm that the final $Z$ distribution was broadly consistent with our goal of having approximately Gaussian-distributed [Fe/H] centred on zero. Indeed, the resulting distribution was [Fe/H]$=0.01\pm0.11$. Our population therefore differs from the applications in \citet{georgyetal2014}, whose model libraries use fixed metallicities, rather than a distribution.

Through down-selection to appropriate mass and metallicity ranges, the grid reduces to 408\,704 models. Each of these models will eventually be given their own rotation rates, inclinations, and so on (Sec.\,\ref{sssec:rotation_intro} onwards), but to ensure that there are always enough models for meaningful population studies, we first triplicated the down-selected grid to 1.23\,million models.

\subsection{Implementing rotation and binarity}
\label{ssec:rotation_and_binarity}

At A and late-B spectral types, rotation and binarity are linked. Close binary systems are tidally braked, with most systems that have periods below 20\,d showing some amount of spin-orbit synchronisation \citep{abtetal2002}. In the absence of a close binary population, the rotational velocity distribution in our mass range of interest consists only of a rapidly rotating population \citep{zorec&royer2012}. However, when close binaries are included, there is also a (somewhat smaller) slowly rotating population. We emulate this by first assigning a boolean binary-star property (without companion properties), then drawing rotational properties. The binary companion mass and other properties are drawn later (Sec.\,\ref{sssec:binary_properties}). We assume that all binaries are unresolved and non-interacting.

We also compute a separate synthetic population without any binaries, which we call the {\tt no\_binaries} sample. This is because some stars, especially exoplanet hosts, are very thoroughly vetted for binary companions, which can sometimes be ruled out through a combination of reconnaissance spectroscopy, high-resolution imaging, and time-series radial velocity observations. Next-generation astrometry from Gaia DR4 will soon add to this list.

\begin{figure}
\begin{center}
\includegraphics[width=0.48\textwidth]{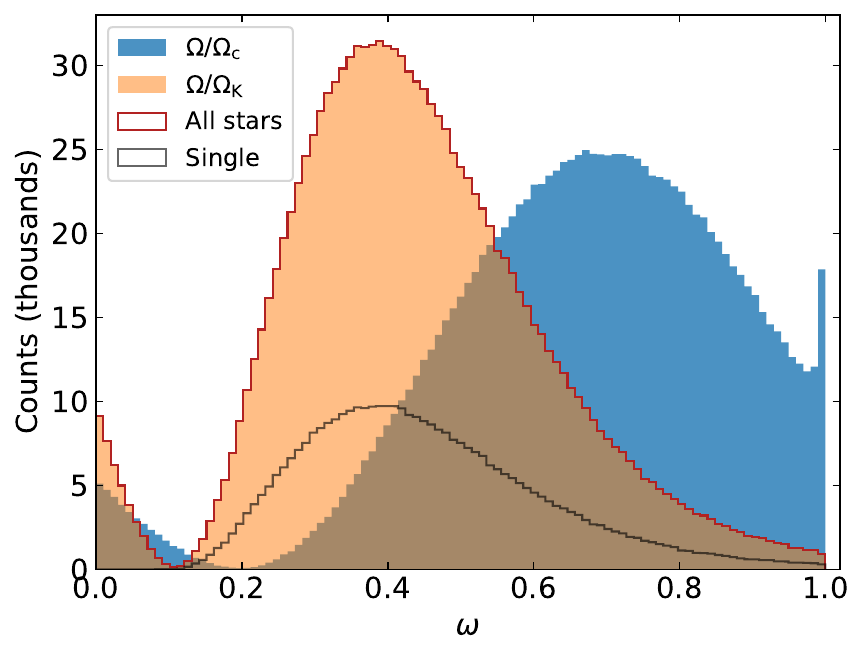}
\caption{Distribution of angular rotation frequencies, $\Omega$, as a fraction of the Keplerian value, $\Omega_{\rm K}$ (orange), and the critical value, $\Omega_{\rm c}$ (blue), in our simulated population of 1.06 million models across 250 bins. 
The red outline encompasses all stars using the $\Omega/\Omega_{\rm K}$ distribution; the black line represents the single stars in the same population. The latter are all fast rotators, while the {\tt full} sample (binaries + single stars) also features a small slowly-rotating component. Details of the Maxwellian distributions sampled to construct this population are given in Sec.\,\ref{sssec:rotation_draw}.}
\label{fig:rotational_distribution}
\end{center}
\end{figure}

If the system is single, including for the special {no\_binaries} sample, the rotation properties come from the rapidly rotating population only, whereas if the system is binary, the rotation velocity is drawn from the full population. In the latter case, the primary still has a somewhat lower chance of being a slow rotator than a rapid rotator, which reproduces the fact that only close binaries -- which comprise only a small subset of the orbital separation distribution -- are tidally braked. We show the results of our emulation in Fig.\,\ref{fig:rotational_distribution}, and will now describe the details.

\subsubsection{Drawing binary stars}
\label{sssec:binary_draw}

Binary stars were simulated by first deciding whether each star has one or more companions with a mass ratio, $q$, above 0.1. The multiplicity fraction for A/F stars (at $q>0.1$) is 0.7 (figures 37 \& 38, \citealt{moe&distefano2017}). We therefore drew a uniform random number between 0 and 1 for each star, and labelled a star as `binary' if that number was below 0.7. This is not strictly the same as the true binary fraction, since in reality some stars have more than one companion, but to allow stars to have multiple companions would markedly increase the complexity and this treatment is intended to be simple -- not a full binary population synthesis model. Hence, while we overestimate the number of stars that have one companion, we ignore the fact that some stars have multiple companions.


\subsubsection{Rotation}
\label{sssec:rotation_intro}

Rotation has a large effect on a star's effective temperature and luminosity due to the centrifugal deformation of the star into an oblate spheroid. The centrifugal acceleration is greatest at the stellar equator, which in the limit of critical rotation exceeds the polar radius by a factor of 3/2 \citep{maeder&meynet2004}. As a consequence, the effective gravity is lower, and the star requires a lower core luminosity in order to counteract gravitational pressure. Thus, a rotating star is less luminous than its non-rotating counterpart.

The dimmer stellar luminosity corresponds to a cooler effective temperature. Even though the surface temperature is a function of latitude -- a topic to which we will return shortly -- one can still talk about the effective temperature of the star as a whole. The definition remains the same, namely,
\begin{eqnarray}
    \label{eq:stefan_boltzmann}
    T_{\rm eff} = \left(\frac{L}{\sigma S}\right)^{\frac{1}{4}},
\end{eqnarray}
where $\sigma$ is the Stefan--Boltzmann constant, but how should one calculate the surface area, $S$, when there is no unique stellar radius? Various options have been implemented in the literature. {\sc mesa} uses the surface-averaged photospheric radius, which is defined as the point at which the optical depth, $\tau$, equals two-thirds. \citet{georgyetal2014} calculated the surface area of the oblate spheroid of known polar and equatorial radius, while a third option is to use the volumetric-equivalent radius $(R_{\rm pol}R_{\rm eq}^2)^{1/3}$. Our approach is functionally the same as the {\sc mesa} one (Sec.\,\ref{sssec:centrifugal}). We show in Appendix\,\ref{app:teff} that the above choices give highly consistent $T_{\rm eff}$ values. The fact that all three options correspond to a greater surface area than a non-rotating, spherical star is another reason that the global $T_{\rm eff}$ is lower.

For a rotating star, surface temperature is a function of latitude. At the equator, where the effective gravity is lowest, the star is substantially cooler than at its poles. An observer viewing the star equator-on will see a cooler, dimmer star. This effect is commonly known as gravity darkening, but the term is problematic because pole-on viewing angles will give brighter luminosities, which some authors call gravity brightening, and it conflates viewing-angle effects with centrifugal effects.

Hence, there are two effects that one must account for, when estimating the effect of rotation on an observed stellar temperature: the global effects of centrifugal deformation (Sec.\,\ref{sssec:centrifugal}), and the geometric effects of viewing angle (Sec.\,\ref{sssec:inclination}).

As previously outlined, our approach is to calculate non-rotating models, then apply the effects of rotation {\it a posteriori}. The advantage of this approach is that it is trivial to implement any distribution of rotation velocities and inclination angles without needing to compute a stellar evolutionary track for every model. The drawback is that the evolutionary effects of rotation are not accounted for, namely, the models' rotation rates do not change as the stars evolve, nor do the models have rotationally-induced mixing of fresh hydrogen into the core, which is known to extend main-sequence lifetimes in a way that is similar to having more core overshooting \citep{georgyetal2013}. The latter process becomes more significant with age, hence its importance to the main-sequence turn-off. However, how much mixing occurs is an ongoing question, and recent work suggests that mixing is `weak' \citep{martinellietal2025}. We therefore acknowledge its absence in our models and do not attempt any ad-hoc correction.

\subsubsection{Drawing rotation velocities}
\label{sssec:rotation_draw}

Our rotational velocities are based on the parameterization of observed distributions of A stars \citep{royeretal2007}. Specifically, we used lagged Maxwellian distributions following \citet{zorec&royer2012}, who provided those distributions depending on stellar mass. Each Maxwellian spanned 0.4\,M$_{\odot}$, beginning with 1.6--2.0, then 1.7--2.1\,M$_{\odot}$, and so on. We used the first five (up to 2.4\,M$_{\odot}$), and took the middle value of each mass range as its representative mass. In each case, we used the distribution closest to the mass of the model. For our stars with $1.4\leq M \leq1.6$\,M$_\odot$, which fall outside of the range in \citet{zorec&royer2012}, we had to create an additional Maxwellian with a slightly smaller mean (104\,km\,s$^{-1}$ rather than 114\,km\,s$^{-1}$) to match the observed distribution in \citet{royeretal2007}.

The lagged Maxwellians' parameters (`scale' = $\alpha$, and `lag' = $\ell$) were applied to draw the equatorial rotation velocity, $v_{\rm eq}$, for that model, using
\begin{eqnarray}
\phi(x) = \sqrt{\frac{2}{\pi}} \frac{(x - \ell)^2}{\alpha^3} e^{-(x - \ell)^2/(2\alpha^2)}
\end{eqnarray}
from \citet{zorec&royer2012}, where $x$ is the velocity range. Each of these Maxwellians are bimodal distributions with a `fast' and `slow' component. For single stars, only the fast component was used, but for binaries, the entire bimodal distribution was used. Every model was assigned an equatorial rotation velocity independently in this manner. We later redraw a small fraction of these velocities, where they are super-critical for the stars they are assigned to (Sec.\,\ref{sssec:redraw}).

\subsubsection{Calculating angular rotation velocities}
\label{sssec:omega}

The calculation of angular rotation velocities, $\Omega$, is complicated because $\Omega$ depends on the equatorial radius, $R_{\rm eq}$, which itself depends on the rotation rate [i.e. $R_{\rm eq}=R_{\rm eq}(\Omega)$] as the star becomes increasingly centrifugally distorted. We describe these calculations in this section.

In this work, we use the Roche approximation in our calculations of the stellar structure as a function of rotation. This approximation asserts that the surface of the star is defined by an equipotential and that this potential arises from a centrally concentrated mass. In other words, the equatorial bulge arising from rapid rotation does not alter the distribution of mass within the star. This assumption has been extensively tested and shown to be accurate to within 1\% error even at critical rotation, and is generally far better \citep{orlov1961,owockietal1994,zahnetal2010,vanbelle2012}. The Roche approximation also implies that the potential at the equator equals the potential at the pole,
\begin{eqnarray}
    -\frac{GM}{R_{\rm p}} = -\frac{GM}{R_{\rm eq}} - \frac12 \Omega^2 R_{\rm eq}^2, \label{eq:roche_potential}
\end{eqnarray}
where $R_{\rm p}$ is the polar radius, which under the Roche approximation is equal to the non-rotating radius and unlike $R_{\rm eq}$ it does not depend on $\Omega$. We note that $\Omega^2 R_{\rm eq}^2 = v_{\rm eq}^2$, which are the linear velocities we obtained in Sec.\,\ref{sssec:rotation_draw}. It follows that
\begin{eqnarray}
    \frac{1}{R_{\rm eq}} = \frac{1}{R_{\rm p}} - \frac12 \frac{v_{\rm eq}^2}{GM}
\end{eqnarray}
and we obtain a relation for the equatorial radius as a function of the polar / non-rotating radius and the equatorial velocity,
\begin{eqnarray}
    R_{\rm eq} = R_{\rm p} \frac{1}{1 - \frac{R_{\rm p} v_{\rm eq}^2}{2GM}}. \label{eq:req}
\end{eqnarray}
This allows angular velocities to be calculated using $\Omega = v_{\rm eq}/R_{\rm eq}$.

It is convenient to express angular velocities as a function of the critical angular velocity, $\Omega_{\rm c}$, or the Keplerian angular velocity, $\Omega_{\rm K}$, but much confusion has proliferated in the literature. We try to give a clear account here and point out potential pitfalls.

The Keplerian angular velocity describes the angular velocity of a body in a circular orbit at the equator, which is
\begin{eqnarray}
    \Omega_{\rm K} = \sqrt{\frac{GM}{R_{\rm eq}^3}}, \label{eq:Omega_K}
\end{eqnarray}
where it is important to remember that $R_{\rm eq}=R_{\rm eq}(\Omega)$, and so $\Omega_{\rm K}$ {\em decreases as the rotation velocity increases}. Substitution into eq.\,\ref{eq:req} and defining $\omega_{\rm k} = \Omega/\Omega_{\rm K}$, we see that
\begin{eqnarray}
    \frac{R_{\rm p}}{R_{\rm eq}} =  1 - \frac{R_{\rm p} R_{\rm eq}^2 \Omega^2}{2GM} = 1 - \frac{R_{\rm p}}{2R_{\rm eq}}\omega_{\rm k}^2,
\end{eqnarray}
leading to the convenient expression for the equatorial radius for any value of $\omega_{\rm k}$:
\begin{eqnarray}
    R_{\rm eq} = R_{\rm p} \left(1 + \frac{1}{2} \omega_{\rm k}^2\right). \label{eq:req_omega}
\end{eqnarray}

There exists a rotation value for each star where it is no longer able to hold onto its equatorial material. 
This is achieved when the effective gravity at the equator is zero,
\begin{eqnarray}
    g_{\rm eff} = \frac{GM}{R_{\rm eq}^2} - \Omega^2R_{\rm eq} = 0.
\end{eqnarray}
There, the angular rotation velocity is equal to the Keplerian angular velocity, i.e. $\omega_{\rm k} = 1$, and from eq.\,\ref{eq:req_omega} we have
\begin{eqnarray}
    R_{\rm eq}\rvert_{\omega_{\rm k}=1} = \frac{3}{2}R_{\rm p},\label{eq:req_crit}
\end{eqnarray}
which is a classical result of Roche theory \citep[e.g.][]{cranmer&owocki1995,zahnetal2010}. From eq.\,\ref{eq:Omega_K} it follows that the critical angular velocity is then defined as
\begin{eqnarray}
    \Omega_{\rm c} = \sqrt{\frac{8 G M}{27 R_{\rm p}^3}}. \label{eq:Omega_c}
\end{eqnarray}
For convenience, we also define $\omega_{\rm c} = \Omega/\Omega_{\rm c}$.

Importantly, $\omega_{\rm k}$ and $\omega_{\rm c}$ are only equivalent at zero and at the point of criticality ($\omega_{\rm k} = \omega_{\rm c} = 1$). At all other times, $\omega_{\rm k} < \omega_{\rm c}$. One must pay careful attention in the literature as to which quantity is being used. The seminal paper on gravity darkening of rapidly rotating models \citep{espinosalara&rieutord2011} is careful with definitions and uses $\omega_{\rm k}$, as does the {\tt GDit} code\footnote{\url{https://github.com/aarondotter/GDit}} that implements their equations. MIST isochrones calculated with rotation \citep{choietal2016} use the {\it notation} $\omega_{\rm c}$ but their {\it formulae} pertain to $\omega_{\rm k}$. The fifth {\sc mesa} paper \citep{paxtonetal2019}, which upgraded the implementation of rotation, refers to `critical rotation' in section 4 but provides the equation for $\Omega_{\rm K}$.\footnote{It is only an issue with notation, the mathematics is fine.} 
Users may work in $\omega_{\rm c}$ or $\omega_{\rm k}$ as preferred, but be aware that they are not interchangable, and specific tools demand specific inputs, such as $\omega_{\rm k}$ for {\tt GDit} and for many of our expressions such as eq.\,\ref{eq:req_omega}. Using the wrong term, or forgetting the $\Omega$-dependence of $R_{\rm eq}$ in eq.\,\ref{eq:req_omega}, are examples of the pitfalls to which we referred earlier.

We showed our distribution of rotation rates in terms of $\omega_{\rm k}$ and $\omega_{\rm c}$ in Fig.\,\ref{fig:rotational_distribution}, where the significance of their differences is evident. 
We have also created a helpful animated gif that illustrates how $\omega_{\rm c}$, $\omega_{\rm k}$, and the Roche geometry change with linear increases in $v_{\rm eq}$ for a specific model.\footnote{\url{https://github.com/gautam-404/roche_model/blob/main/rotation_roche.gif}} Hereafter, we only use $\omega_{\rm k}$ unless making a direct comparison, and to further alleviate confusion we always use the subscripts.


\subsubsection{Redrawing velocities for super-critical rotators}
\label{sssec:redraw}

When drawing rotational velocities (Sec.\,\ref{sssec:rotation_draw}), a small fraction of stars ($\sim$1.11\,\%) were assigned velocities that exceeded those permitted by the Roche model for their mass and radius. In such cases, the implied rotation is super-critical, in the sense that the corresponding value of $\omega_{\rm k}$ exceeds unity, and the model cannot be supported in hydrostatic equilibrium. These draws are unphysical for the sampled stellar model and must be rejected. We classified draws with $\omega_{\rm k} > 0.999$ as super-critical and redrew them.

Redrawing the rotational velocity rather than discarding the star is necessary because the maximum physically allowed rotation depends on the stellar properties, i.e. a velocity that is super-critical for one star may remain sub-critical for another with a different mass and radius. Simply removing all stars above some fixed velocity threshold would therefore artificially suppress the high-rotation tail of the population, and bias other stellar properties such as radius (and indirectly, age).

The redraws were performed using the same lagged Maxwellian prescription described in Sec.\,\ref{sssec:rotation_draw}, but with the probability density truncated so that only the high-rotation part is retained. We adopted this truncated redraw to limit the bias of the resampled stars toward systematically lower rotation. Specifically, we restricted each mass dependent distribution in Sec.\,\ref{sssec:rotation_draw} to the interval between the mode and the maximum velocity, and renormalised the resulting probability density function before redrawing. The truncated distribution is then
\begin{eqnarray}
    p_{\rm redraw}(x) =
    \begin{cases}
        \dfrac{\phi(x)}{\int_{x_{\rm mode}}^{x_{\rm max}} \phi(x')\,\mathrm{d}x'}, & x_{\rm mode} \le x \le x_{\rm max}, \\
        0, & \text{otherwise},
    \end{cases}
\end{eqnarray}
where $f(x)$ is the lagged Maxwellian defined in Sec.\,\ref{sssec:rotation_draw}, $x_{\rm mode}=\ell+\sqrt{2}\,\alpha$, and $x_{\rm max}$ is the upper limit of the adopted velocity range.

This allowed us to preserve the rapidly rotating part of the adopted observed distribution for these stars while making sure that the final assigned velocity was physically realisable for each synthesised model. If a redraw remained super-critical, the process was repeated until a sub-critical velocity was obtained.

\subsubsection{Calculation of $\log L$ and global $T_{\rm eff}$ for rotating models}
\label{sssec:centrifugal}


To account for the effects of centrifugal deformation on the observables of our models, we first calculate the luminosity by reference to a grid of rotating models computed with {\sc mesa} using the same input physics as \citet{gautametal2026}. The reference grid was evaluated at a fixed age of 40\,Myr and metallicity $Z=0.015$, while spanning the full mass range of our population synthesis models, from 1.4 to 2.5\,M$_\odot$, and a broad range of rotation rates. All models in this reference grid have uniform radial rotation profiles, i.e. ${\rm d}\Omega/{\rm d}r = 0$.

For each model in this reference grid, we calculated the ratio of the log-luminosity of the rotating model to that of the corresponding non-rotating model, $\log L /{\log L}_0$, as a function of stellar mass, $m$, and $\omega_{\rm K}$. We then constructed a 2-D interpolation function in $(m,\omega_{\rm k})$, which allows the luminosity of a rotating model to be obtained from its non-rotating counterpart.


Having determined the luminosity of the rotating model from our interpolation function, we then calculated its global effective temperature from the Stefan--Boltzmann relation (eq. \ref{eq:stefan_boltzmann}) and the surface area, $S$, of the centrifugally deformed star. For the surface area, we adopt the Roche-surface approximation used in \citet{paxtonetal2019},
\begin{eqnarray}
    S_\mathrm{Roche} = 4\pi R_\mathrm{eq}^2 \left(1 - \frac{\omega_{\rm k}^2}{3} + 0.08525\,\omega_{\rm k}^4 - 0.04908\,\omega_{\rm k}^6 \right), 
\end{eqnarray}
where $R_{\rm eq}$ is the equatorial radius. We refer the reader to Appendix \ref{app:teff} for a comparison with other shape-based methodologies for calculating the effective temperature of a deformed star.

\begin{figure*}
\begin{center}
\includegraphics[width=0.98\textwidth]{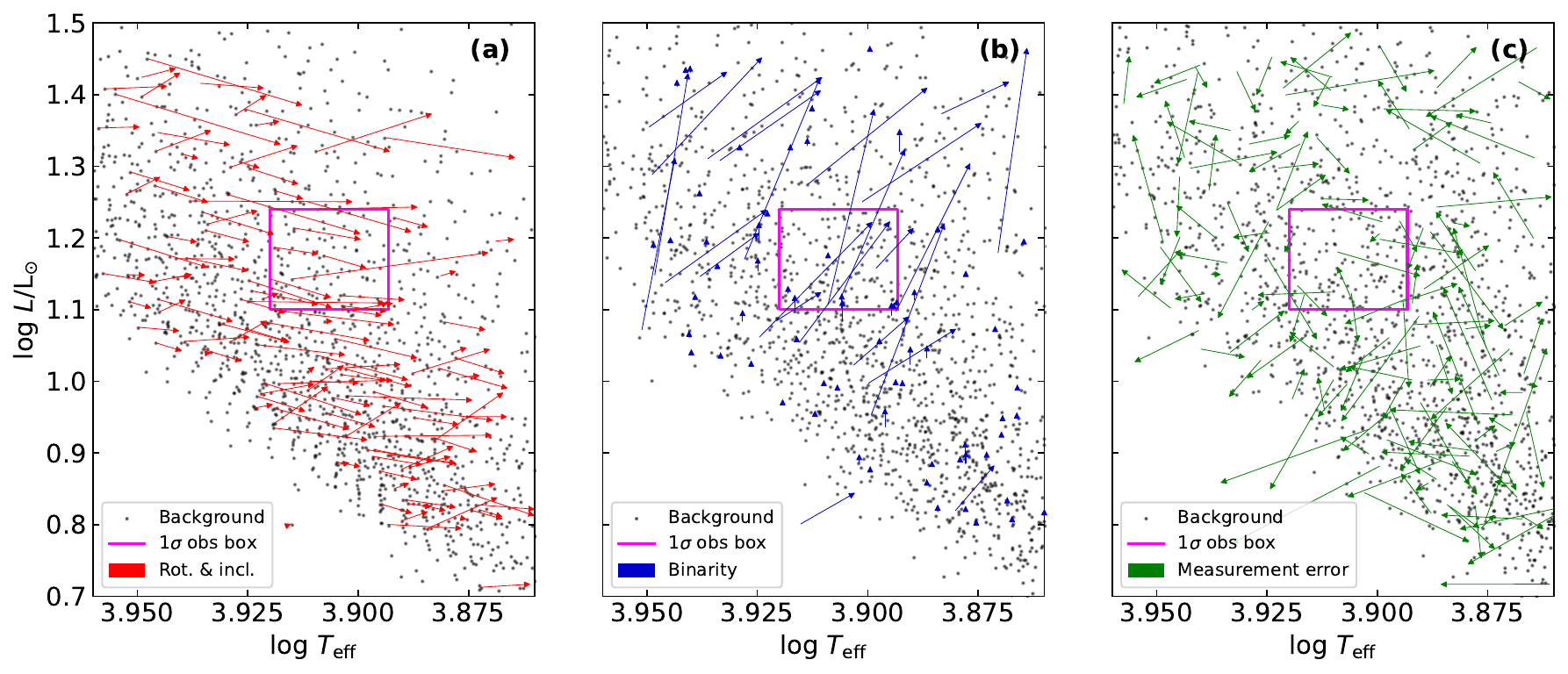}
\caption{Individual additions to the scatter of stars in the HR diagram relative to single, non-rotating stellar models. {\bf (a)} The combined effects of rotation (centrifugal acceleration) and inclination (viewing angle). {\bf (b)} The effect of any binary companion. {\bf (c)} Random measurement uncertainty, or `observational scatter'. The small grey points show the background distribution of simulated stars, thinned by a factor 250. Each panel shows an observational box (magenta), whose edges are $\pm$1$\sigma$ from its centre, using typically-adopted observational uncertainties (defined in Sec.\,\ref{ssec:obs_uncertainty}), 
i.e. $\pm250$\,K in $T_{\rm eff}$ (0.0268 in $\log T_{\rm eff}$) and $\pm$0.07\,dex in $\log L$.
Each arrow traces the same simulated star as each effect is applied, initially thinned by a further factor of 5, but only plotted if both ends of the arrow are inside the plot window, hence there are more arrows in the first panel. Arrows for binarity are only drawn if the simulated star is a binary, causing a further reduction in the number. The final panel gives the impression that random scatter is biased to the lower-left, but this is just the combined result of only drawing arrows if both ends are inside the plot window and the stars having moved rightward in the preceding panels. The combined effects are summarised in Sec.\,\ref{ssec:summary}.}
\label{fig:scatter_contributions}
\end{center}
\end{figure*}

We found that the fractional $T_{\rm eff}$ reduction was almost independent of stellar mass, peaking at 0.127 at $\omega_k$ = 0.999. The luminosity reduction is not a linear function of mass, and lower-mass stars are most affected. The fractional reductions in $\log L$ at the maximum rotation rate were $0.034$ for $M=1.4$\,M$_{\odot}$ and $0.014$ for $M=2.5$\,M$_{\odot}$. Since the magnitude of the reduction in $T_{\rm eff}$ is much greater than that of $\log L$, the net effect in the HR diagram is to shift stars to the right and slightly down (Fig.\,\ref{fig:scatter_contributions}a). This is consistent with our expectations set out in Sec.\,\ref{sssec:rotation_intro}.

For most of the main-sequence evolution, we do not expect age to have an appreciable effect on this method. The Roche approximation assumes a centrally condensed star, and the degree of central condensation does not change much on the main sequence. Indeed, \citet{espinosalara&rieutord2011} found negligible difference in the latitudinal dependence of $T_{\rm eff}$ when changing the hydrogen abundance in the core. Nonetheless, we checked the validity of our fixed-age interpolation by performing a multi-dimensional interpolation (on the same rotating models) that also included variable age up to the TAMS. For models up to 90\% of the TAMS age, the standard deviation of the $T_{\rm eff}$ differences between the two approaches is 14.1\,K. When limiting to 2/3 of the TAMS age, this drops to 6.2\,K. These values pale in comparison to the 250\,K of random measurement uncertainty applied later (Sec.\,\ref{ssec:obs_uncertainty}). However, at the TAMS, whose location occurs at different age for stars of different mass, an interpolator that has both $\omega_{\rm k}$ and age as variables does not perform well. This is why we fixed the age at the early main sequence for our interpolation.

\subsubsection{Viewing angle (inclination) effects}
\label{sssec:inclination}

To account for the geometric effect of viewing angle, we drew a random inclination angle for each star. For an isotropic distribution, the probability of observing a star at some inclination, $i$, is proportional to $\sin i$. Thus, we assigned each star an inclination following $P(i) \propto \sin i$. When combined with the rotation rate, the inclination determines the amount of observed gravity darkening (or brightening). In the Roche model approximation, the shape of the stellar surface depends only on $\omega_{\rm k}$, and the geometric correction due to viewing angle thus depends only on $i$ and $\omega_{\rm k}$.\footnote{One could, in principle, use $\omega_{\rm c}$ instead, but since $\omega_{\rm k} \neq \omega_{\rm c}$, the functional dependence is different. Further, {\tt GDit} returns values based on inputs of $\omega_{\rm k}$, hence we refer explicitly to $\omega_{\rm k}$, here.} We used pre-computed geometric correction factors $C_T$ and $C_L$ calculated with {\tt GDit} and packaged with {\sc mesa} for interpolation and repeated use. This is much faster than $10^6$ function calls. We determined the inclination of nil effect, i.e., where the apparent $T_{\rm eff}$ and $L$ are equal to their globally averaged quantities, to be 55$^{\circ}$.

The geometric effect adds 63\,K and 1.5\,L$_{\odot}$ of scatter to the temperature and luminosity distributions of the stars, respectively. There was only a small systematic shift in these distributions by 3\,K cooler and 0.0024\,L$_{\odot}$ brighter, respectively. We did not make additional corrections for limb darkening, which \citet{georgyetal2014} showed to be negligible for stars with radiative envelopes in all but the most extreme cases.

Across our 1.22\,million models, the median effect of rotation on the {\it observed} $T_{\rm eff}$ is $-250.9$\,K. The most extreme case has $\Delta T_{\rm eff} = -1477.3$\,K, which occurs for a 2.40-M$_{\odot}$ star rotating at $\omega_{\rm k} = 0.954$ and viewed at an inclination of $i=86.9^{\circ}$. At the 90th percentile, $\Delta T_{\rm eff} = -580$\,K.

\subsubsection{Effect of rotation on density and pulsation mode frequencies}
\label{sssec:delta_density}

Many stars at spectral type A pulsate as $\delta$\,Sct stars \citep{kurtz2022}. Their pulsations are acoustic modes whose frequencies depend on stellar density. We foresee that our synthetic population will be useful in studying populations of pulsators, hence we consider here the effect of centrifugal deformation on mean stellar density, and hence on mode frequencies.

At the mode of the rotational velocity distribution, centrifugal deformation reduces mean stellar density by several percent \citep{murphyetal2022}. As $\Omega_{\rm rot}$ approaches $\Omega_{\rm K}$ (i.e. as $\omega_{\rm k} \to 1$), the density of the oblate spheroid relative to its non-rotating equivalent, $\rho/\rho_{\rm 0}$, approaches 4/9 (= 0.44; derived below). Since the frequencies of acoustic pulsation modes scale with $\sqrt\rho$ \citep{ulrich1986,aertsetal2010}, these are substantially reduced, too. \citet{reeseetal2008} and \citet{garciahernandezetal2015} have demonstrated that the asteroseismic large spacing, $\Delta\nu$, continues to scale approximately with $\sqrt{\rho}$, even at rapid rotation. 

Since the total stellar mass is constant, the effect of rotation on stellar density is governed by the change in the Roche volume, which follows from the scaling $V \sim R_{\rm p}R_{\rm eq}^2$. 
Given $R_{\rm p}=R_0$, the ratio of the new to the old volume is 
\begin{eqnarray}
\frac{V}{V_0} = \frac{R_{\rm p}R_{\rm eq}^2}{R_0^3} = \left(\frac{R_{\rm eq}}{R_0}\right)^2, 
\end{eqnarray}
and it follows from eq.\,\ref{eq:req_omega} that 
\begin{eqnarray}
\frac{\rho}{\rho_0} = \left(\frac{R_0}{R_{\rm eq}}\right)^2 = \left(1 + \frac{1}{2} \omega_{\rm k}^2\right)^{-2}. \label{eq:density}
\end{eqnarray}
With the relation $R_{\rm eq} = (3/2)R_{\rm p}$ at $\omega_{\rm k}=1$ (eq.\,\ref{eq:req_crit}), one readily obtains the earlier statement that $\rho/\rho_0 \to 4/9$ as $\omega_{\rm k} \to 1$. 

We consider all acoustic mode frequencies, $f$, to scale with $\sqrt\rho$, hence they are related to their non-rotating equivalents, $f_0$, via
\begin{eqnarray}
    \frac{f}{f_0} = \left(1 + \frac{1}{2} \omega_{\rm k}^2\right)^{-1}.
\end{eqnarray}
All mode frequencies provided in this work are corrected for this effect.

We only provide frequencies for radial modes, whose frequencies in non-rotating stars were calculated with {\sc gyre} \citep{townsend&teitler2013} as described by \citet{gautametal2026}.
For radial modes, the first-order correction due to rotation in the Ledoux formula is zero \citep{ledoux1951,aertsetal2010}. 
We do not account for second-order or higher corrections \citep[see, e.g.,][]{suarezetal2006b,guoetal2024}.
Additional known effects, such as the focusing of pulsation modes to the equatorial regions of rapid rotators \citep{reeseetal2013}, are also not accounted for here (nor in {\sc gyre}). Our frequencies are intended to be indicative, i.e. valid at the population level, or `approximate'; they should not be used for asteroseismic modelling. Our focus on radial modes also avoids the issue of avoided crossings \citep{aerts&tkachenko2024,gautametal2026}.

We do not demonstrate any applications of mode frequencies in this work, but in the future we intend to use them for population-level inference of pulsation properties, and we provide them for community use. More information is given in Sec.\,\ref{ssec:applet} and the Data Availability Statement.

\subsubsection{Binary star properties ({\tt full} sample only)}
\label{sssec:binary_properties}

\begin{figure}
\begin{center}
\includegraphics[width=0.48\textwidth]{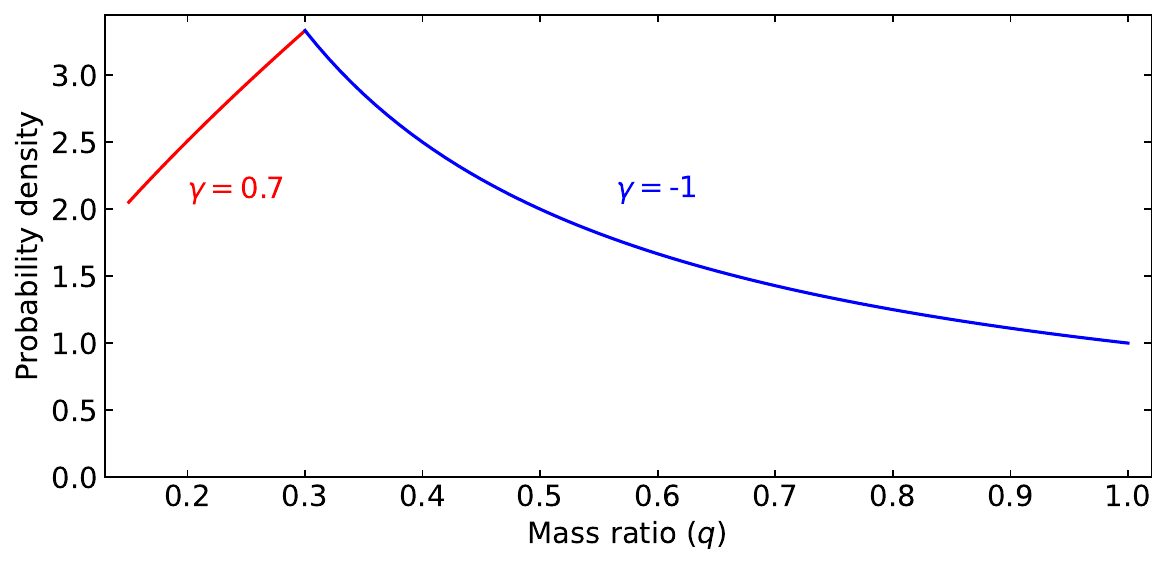}
\caption{Unnormalised probability density for the binary mass ratios.}
\label{fig:binary_probability}
\end{center}
\end{figure}

When drawing companion masses, it is not appropriate to draw from a stellar IMF because companion masses are correlated with the primary mass (and with orbital period). For each binary system we drew a mass ratio according to a segmented power law as formulated by \citet{moe&distefano2017}, that is, we drew from a probability distribution $p_q \propto q^{\gamma}$ with slopes
\begin{eqnarray}
\gamma = & \gamma_{{\rm small}q} \quad {\rm for}  \quad 0.15 \leq q < 0.30,\\
		& \gamma_{{\rm large}q} \quad {\rm for}  \quad 0.30 \leq q \leq 1.00,
\end{eqnarray}
where the mass-ratio probability distribution $p_q$ is continuous at $q=0.3$. Mass ratios are generally smaller at larger orbital separations and the quality with which they are measured in the literature varies markedly according to the sensitivity of various binary detection techniques. The most precise measurement of $\gamma_{{\rm small}q}$ and $\gamma_{{\rm large}q}$ for A-type stars comes from a pulsation-timing study of 341 systems \citep{murphyetal2018}, but this covers only 1\,dex in $\log P_{\rm orb, d}$, whereas binaries are found over approximately 7\,dex. Conversely, the broadest survey covering the same mass range is by \citet{gulliksonetal2016}, covering 3.6\,dex in $\log P_{\rm orb, d}$ but containing an order of magnitude fewer binaries than the pulsation-timing sample (after sample cleaning performed by \citealt{moe&distefano2017}), hence the uncertainties are somewhat larger. The \citeauthor{gulliksonetal2016} sample also only extends to $q=0.75$, but \citet{murphyetal2018} have shown that $p_q$ flattens between 0.75 and 1.0, so the \citeauthor{gulliksonetal2016} result can be trivially extended to $q=1.0$. The \citeauthor{gulliksonetal2016} sample covers the middle of the $\log P_{\rm orb}$ range, and its extended cumulative distribution function ($F_{q,0.1-1.0}$) is almost identical to the \citeauthor{murphyetal2018} sample. Hence, we adopted the \citeauthor{gulliksonetal2016} result of $\gamma_{{\rm small}q}=0.7$ and $\gamma_{{\rm large}q}=-1.0$ (Fig.\,\ref{fig:binary_probability}). The mass ratio is then drawn by evaluating a uniform random number between 0 and 1 against the corresponding cumulative distribution function.

Once a mass ratio is drawn for the system, the companion's luminosity is estimated using a mass--luminosity relation, $L\propto M^{4.328}$ \citep{ekeretal2015}, and the primary mass and luminosity. The luminosity of the two stars is then summed. Since we assume the companions to be lower in mass and the primaries to be on or near the main sequence, the secondaries are also on the main sequence (they are not red giants), and the use of a mass--luminosity relation is justified. This approach avoids the obvious problem that the secondary star will often have a mass lower than the low-mass limit of our underlying model grid. 

\begin{figure}
\begin{center}
\includegraphics[width=0.48\textwidth]{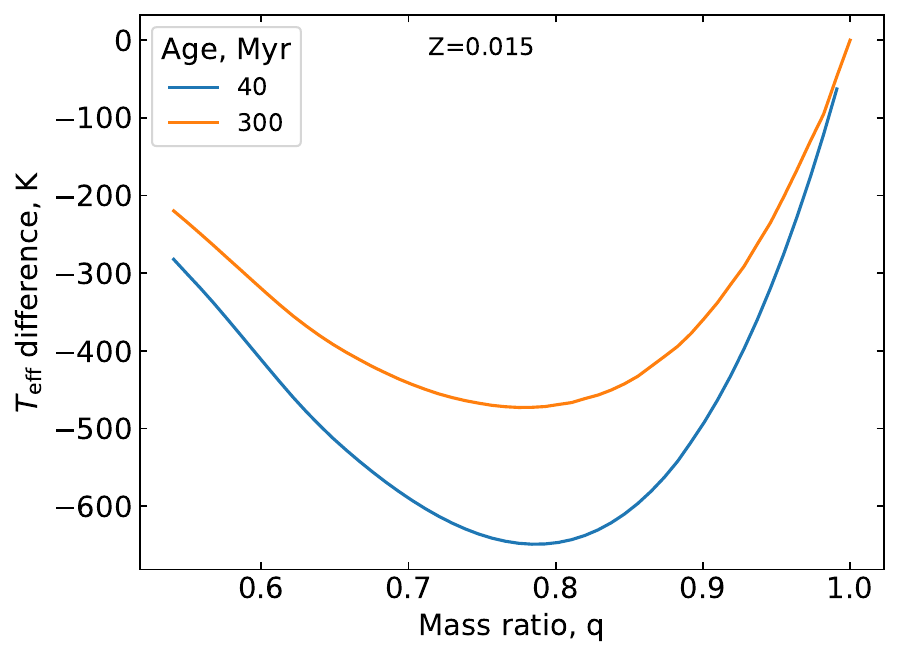}
\caption{Simulated $T_{\rm eff}$ difference between a 2.2-M$_{\odot}$ star observed in isolation and as part of a binary. The binary mass ratio is $q = M_2/M_1$. Because more massive stars evolve more quickly, the amplitude of the difference is age dependent. There is negligible metallicity dependence for our synthetic population.}
\label{fig:teff_diff}
\end{center}
\end{figure}

Estimating the companion's effect on the observed effective temperature is more challenging. Similar problems in the literature have been solved by using fluxes and colours \citep{albrow&ulusele2022}, but these were not computed for our {\sc mesa} model outputs. Simulations in appendix A of \citet{murphyetal2015a} used stellar spectral synthesis and Balmer-line fitting to show that $T_{\rm eff}$ misestimations of 300\,K are `easily introduced' when a binary consisting of two late-A stars are misinterpreted as one star. However, spectral synthesis for each binary is beyond the scope of this work; a simpler approach is required. We simulated the effect on observed $T_{\rm eff}$ values by taking a weighted average of the $T_{\rm eff}$ values of each component, weighting by the luminosity ratio. Since the $T_{\rm eff}$ values are not known for the secondary in our binary synthesis, we sought a generalised form. Specifically, we took pairs of non-rotating stars of the same age and metallicity, and plotted their weighted average $T_{\rm eff}$ as a function of their mass ratio. To access lower mass ratios (and for this purpose only), we calculated extra models using the same {\sc mesa} inlist as \citet{gautametal2026}, down to $M=1.2$\,M$_{\odot}$. 
We found $T_{\rm eff}$ differences peaking at $0.75<q<0.80$ that reach $-600\pm200$\,K, with the uncertainty dominated by stellar age (Fig.\,\ref{fig:teff_diff}). Naturally, at $q=1.0$ the inferred $T_{\rm eff}$ difference is 0. For simplicity, we approximated this in quadratic form as $\Delta T_{\rm eff} = 12000(q-0.8)^2 - 500$ for $0.6 \leq q \leq 1.0$, and set $\Delta T_{\rm eff} = 0$ for $q<0.6$. 
The latter choice is justified because components with disparate temperatures become easy to distinguish in spectral energy distribution (SED) fitting \citep{jadhavetal2021}, hence, would show up in a careful analysis of any given target.
Since the majority of the companions to A stars will have masses that put them below the Kraft break (below $\sim1.4$\,M$_{\odot}$; \citealt{beyer&white2024}), we assumed the secondaries were non-rotating and we did not centrifugally distort them. Our implementation of binarity differs from \citet{georgyetal2014}, who drew companion masses from a uniform random distribution between 0.1 and 1.0, and who adopted the primary $T_{\rm eff}$ as the $T_{\rm eff}$ of the binary.

To evaluate the reliability of our approximation, we considered the eclipsing binary HD\,23642 in the Pleiades (age $\sim$120--170\,Myr), whose component masses and temperatures were measured by \citet{southworthetal2023}. The stars have masses of 2.273 and 1.595\,M$_{\odot}$, hence a mass ratio of $q = 0.70$, and $T_{\rm eff}$ values of 10200 and 7670\,K, respectively. The TIC v8.2 gives $T_{\rm eff}=9962$\,K \citep{paegertetal2021}. Our approximation estimates a $T_{\rm eff}$ reduction of 380\,K, while the actual reduction is ($10200 - 9962=$) 238\,K. The difference of 142\,K is within the typical TIC $T_{\rm eff}$ uncertainty (and well within the actual uncertainty of 449\,K for this target), and the proximity of the two stars in their 2.4-d orbit has undoubtedly affected the $T_{\rm eff}$ of each star through mutual irradiation in this case. In summary, the approximation appears valid for our purposes.

The effect of binaries will always be to {\it increase} the observed luminosity and, if the companion mass is large enough to significantly affect the SED, to {\it decrease} the observed $T_{\rm eff}$ (Fig.\,\ref{fig:scatter_contributions}b). 

\subsection{Observational uncertainty}
\label{ssec:obs_uncertainty}

\subsubsection{When should uncertainty be added?}
When stellar models are used to simulate an observed population, it is important to add measurement uncertainty to the stellar properties. We discuss the magnitude of that uncertainty in this subsection. However, if the model population is used as a reference sample to infer the properties of a real star with its own measurement uncertainties (as in the examples we give later in Sec.\,\ref{ssec:hd250208} and \ref{ssec:hd56414}), then one should not add measurement uncertainty to the models.

\subsubsection{$T_{\rm eff}$ uncertainty}
Observational uncertainties are often poorly quantified, especially in effective temperature. It is common practice to quote random uncertainties resulting from spectroscopic analysis by one specific code with little consideration for the underlying systematics, such as uncertainties arising from inadequacies of model atmospheres, differences in outputs from different spectroscopic analysis codes applied to the same data, or the accuracy to which those codes are calibrated \citep{byrne&stanway2023}. A good summary of this was made by \citet{blanco-cuaresma2019}.  The Gaia General Stellar Parametriser \citep[GSP;][]{recio-blancoetal2023} is a prime example, sometimes returning implausibly small uncertainties of $<10$\,K. The GSP team is aware of the issue.

The same can be said of spectral energy distribution (SED) fitting: in publishing their new {\tt PySSED} code for determining stellar effective temperatures with SEDs, \citet{mcdonaldetal2024} provide a good discussion of the effect of uncertainties on the inputs, but remark that it is not possible to accurately quantify uncertainties on the resulting effective temperatures. It is also known that photometric and spectroscopic methods can give systematically different values \citep{frebeletal2013}.

There are some fundamental calibrators of effective temperature, such as eclipsing binaries \citep{milleretal2020,milleretal2022}, the infrared flux method \citep{casagrandeetal2010,casagrandeetal2021}, and interferometry \citep{maestroetal2013,casagrandeetal2014,jonesetal2015,whiteetal2018}. Alarmingly, indirect methods of inferring temperature are based on only a handful of such calibrator stars or systems, leading to a ``calibration pyramid'' susceptible to unknown systematic errors \citep{tayaretal2022}. Interferometric calibration suggests that spectroscopy is accurate to about 2\% for G stars \citep{whiteetal2018} and to 3\% for B stars \citep{maestroetal2013}, with A stars being somewhere in the middle. But rarely is any underlying accuracy from calibrations of the $T_{\rm eff}$ scale factored in to quoted uncertainties for individual stars. It is then left to individual users to implement a floor in uncertainty (e.g. 2\% in the $T_{\rm eff}$ of A stars; \citealt{murphyetal2020b}).

We considered that one approach to inferring meaningful uncertainties was to inspect the scatter in $T_{\rm eff}$ values from various methods applied to the same star(s). \citet{niemczuraetal2015,niemczuraetal2017} conducted such an analysis for spectroscopic temperatures of A stars determined by different approaches (fitting Balmer lines, fitting Fe lines) and compared them with temperatures from photometry: SEDs and the Kepler stellar properties catalogue of \citet{huberetal2014}. They found typical scatter between spectroscopy and SED-fitting of $\pm200$\,K ($\sim$2.5\%), which is greater than the formal uncertainties quoted (e.g. 100\,K for the Balmer lines and Fe lines analysis). In a more recent example, \citet{kuess&paunzen2025} compared the $T_{\rm eff}$ values of chemically normal A stars from multiple different catalogues of stellar properties, including the Gaia DR3 Apsis catalogue \citep{fouesneauetal2023} and the revised TESS Input Catalogue \citep{stassunetal2019}. Between those two catalogues, they found a mean $T_{\rm eff}$ difference of 455\,K and a standard deviation of 663\,K. This large difference is alarming, and far exceeds the typical $T_{\rm eff}$ uncertainties of A stars quoted in the TIC, which average a little under 200\,K. \citet{kuess&paunzen2025} also compared two other catalogues in permutation with the TIC and Gaia Apsis pipelines, and found differences averaging $\sim$200\,K.

In light of the above, namely a floor of at least 2\% uncertainty and characteristic scatter of $\sim$200\,K, it seems that a 250-K uncertainty (those values added in quadrature) is a reasonable choice for late A stars.

\subsubsection{Luminosity uncertainty}
For luminosity, the uncertainty estimate is much more situational. In most cases, the uncertainty is dominated by two sources, namely parallax and extinction, both increasing with distance. Extinction is highly variable across the sky, even at short distances and away from the galactic plane \citep{greenetal2019}. This can make the uncertainty correspondingly variable. Extinction is also filter dependent, meaning the uncertainty on bolometric correction should be accounted for, and \citet{tayaretal2022} pointed out that bolometric fluxes (and hence, luminosities) are accurate to a fundamental floor of 2.4\%\,$\pm$\,0.6\%. Parallax uncertainties are shrinking with successive Gaia data releases \citep{gaiacollaboration2023a}, and for most TESS objects of interest or asteroseismic targets, parallax is no longer the dominant contributor to the luminosity uncertainty. This might not hold true for upcoming space telescopes with larger mirrors (such as Roman; \citealt{weissetal2025}) targeting farther stars.

This situational nature of the uncertainty budget makes it difficult to elucidate a single value. There are also too few hot exoplanet hosts to infer a `typical' uncertainty used in the literature, so instead we have sought one from asteroseismic work. For the cluster NGC\,2516, which lies some 1.3\,kpc distant and includes many $\gamma$\,Dor and $\delta$\,Sct stars, \citet{glietal2024} adopted the equivalent of 0.13\,dex of uncertainty in $\log L$. In the study of the Cep--Her Complex, which lies somewhat closer at $340\pm30$\,pc, \citet{murphyetal2024} provided uncertainties equating to $\sigma(\log L) \approx 0.04$\,dex. Outside of clusters, the values are less homogeneous: for \textit{Kepler} $\delta$\,Sct stars, \citet{balona2018c} used 0.052\,dex, which was provided without discussion; \citet{murphyetal2019} implemented Monte Carlo simulations that yielded a median uncertainty of $\sigma(\log L) = 0.029$\,dex; and \citet{bowman&kurtz2018} used $\log g$ in place of $\log L$. Other recent asteroseismic studies of TESS A stars have assumed an uncertainty \citep[0.1\,dex;][]{durfeldt-pedrosetal2024}, or did not calculate one \citep{gootkinetal2024}.

The above discussion highlights how variable luminosity uncertainties can be. For this work, we ultimately adopted a simple average of the above five numbers for 0.07\,dex of $\log L$ uncertainty, noting that our motives spanned both the calculation of realistic uncertainties where possible and capturing those used by others.

\subsubsection{Summary of observational uncertainty}
In summary, we added Gaussian-distributed observational uncertainty to the $T_{\rm eff}$ (250\,K) and $\log L/{\rm L}_{\odot}$ (0.07\,dex) values, providing these in additional columns of our output files. These values correspond to the observational box in Fig.\,\ref{fig:scatter_contributions}. The magnitude of the uncertainties above may need modification when applied to certain stellar populations, e.g., nearby stars with smaller parallax uncertainties. Users may wish to replicate our detailed recipe to produce synthetic populations suited to their use cases. For further reading, \citet{tayaretal2022} provide an in-depth analysis focused on Sun-like stars.

\begin{figure}
\begin{center}
\includegraphics[width=0.48\textwidth]{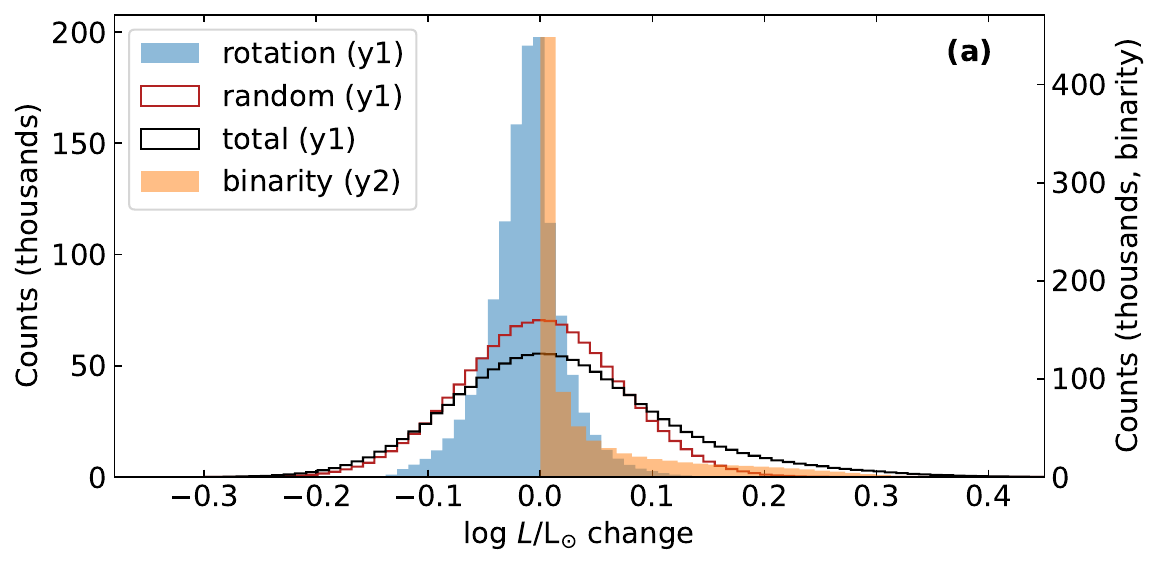}
\includegraphics[width=0.48\textwidth]{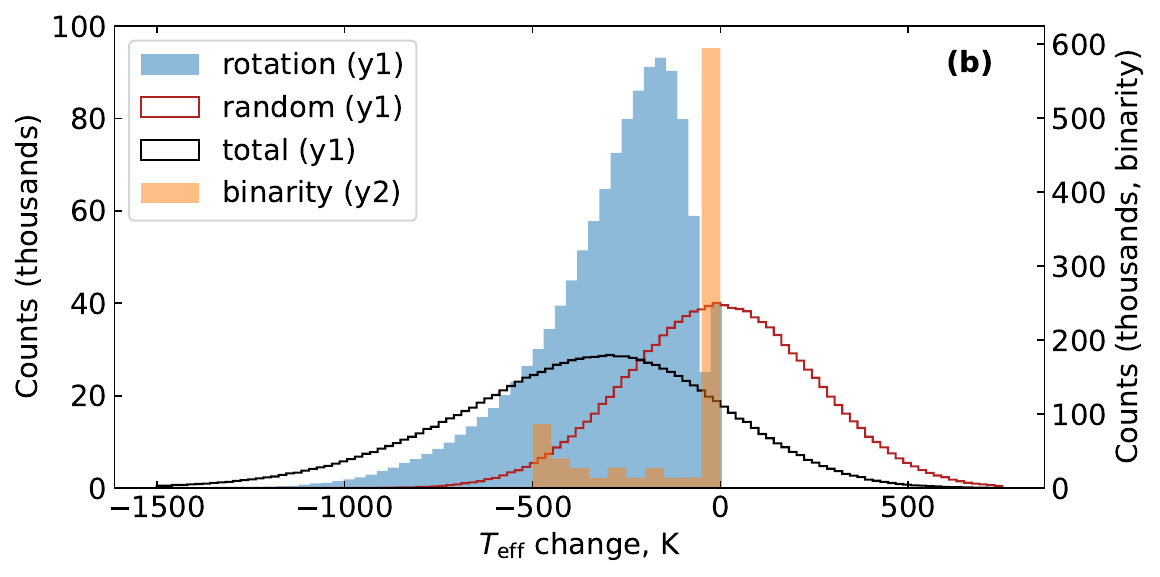}
\caption{{\bf (a):} The distribution of $\log L/{\rm L}_{\odot}$ differences, compared to single non-rotating stars, caused by binarity (right `y2' axis), by rotation (left `y1' axis), and by simulated random measurement error (left axis). Also shown is the total (sum) of these distributions (left axis). {\bf (b):} As above, but in $T_{\rm eff}$. {\bf Notes:} For the binarity histograms, only the values for binary stars are shown, otherwise non-binaries contribute an even larger peak at $x=0$ in both panels. Although measurement error is random, binaries only {\it add} to the luminosity and {\it subtract} from the inferred $T_{\rm eff}$. Similarly, the effect of rotation is always to decrease the observed $T_{\rm eff}$.}
\label{fig:teff_and_logl}
\end{center}
\end{figure}

\subsection{Summary of population synthesis}
\label{ssec:summary}

We have taken a grid of models of `intermediate' mass (1.4--2.5\,M$_{\odot}$) and sampled them with a Salpeter-like IMF. Their metallicities were sampled in accordance with young A stars in the solar neighbourhood, and we ensured that models were sampled equitably in age so as not to bias any particular evolutionary phase. We randomly drew for stars to be binaries according to known binary fractions, and drew companion masses consistent with measured mass-ratio distributions for A stars. We also drew rotational velocities from the measured distributions for A stars, accounting for dependence on both stellar mass and binarity. The centrifugal effects of rotation were applied. Each star was assigned a viewing angle (inclination), which further affects the observed properties. Random observational error was also added. The various contributions to the final temperatures and luminosities, and their differences from single non-rotating models, are shown in Fig.\,\ref{fig:teff_and_logl}.

Hereafter we distinguish between five stages of stellar parameters, which are all available as separate columns in the output files (see also Fig.\,\ref{fig:overview}):
\begin{enumerate}
    \item {\bf Original} properties are those of the original non-rotating models, without any effects of rotation, binarity, or random measurement uncertainty applied.
    \item {\bf Model} properties are those of rotating model stars, after centrifugal deformation has been applied, but without any effects specific to the observer, such as viewing angle.
    \item {\bf Inclined} properties are the model stars (ii), further modified to apply viewing-angle effects arising from the inclination and rotation rate.
    \item {\bf Binary} properties are the inclined properties (iii) after also applying the $T_{\rm eff}$ and $\log L$ effects of a binary companion. This constitutes a reference stellar population against which to compare real observed stars. This stage is skipped for the special {\tt no\_binaries} sample.
    \item {\bf Observed} properties correspond to the previous step (iv in the {\tt full} sample, or iii in the special {\tt no\_binaries} sample), plus random observational error, useful for theoretical population-level studies. By separating this column from its previous steps, it becomes trivial for users to apply measurement uncertainty of a different magnitude.
\end{enumerate}

In the following applications, we explicitly state to which properties we refer and help the reader understand which properties to use in various contexts.

\section{Application to stellar parameter inference}
\label{sec:applications}
\subsection{Demonstration that age inference can be biased}

\begin{figure}
\begin{center}
\includegraphics[width=0.48\textwidth]{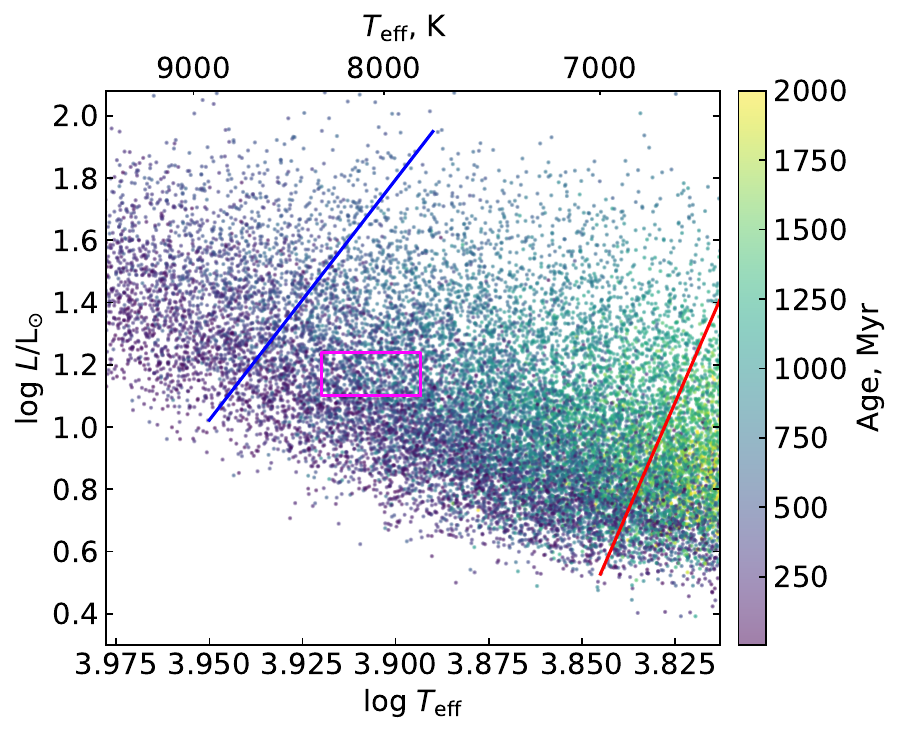}
\caption{Hertzsprung--Russell diagram of the synthesised population, thinned to every 50th star, colour-coded by age. Solid blue and red lines are the theoretical $\delta$\,Sct instability strip edges determined with time-dependent convection models \citep{dupretetal2004}. {\it Observed} properties are shown. The magenta box is used to show the scattering effects summarised in Sec.\,\ref{ssec:summary}.}
\label{fig:HRD}
\end{center}
\end{figure}

An HR diagram of the {\it observed} parameters of our synthesised population is shown in Fig.\,\ref{fig:HRD}.

\begin{figure}
\begin{center}
\includegraphics[width=0.48\textwidth]{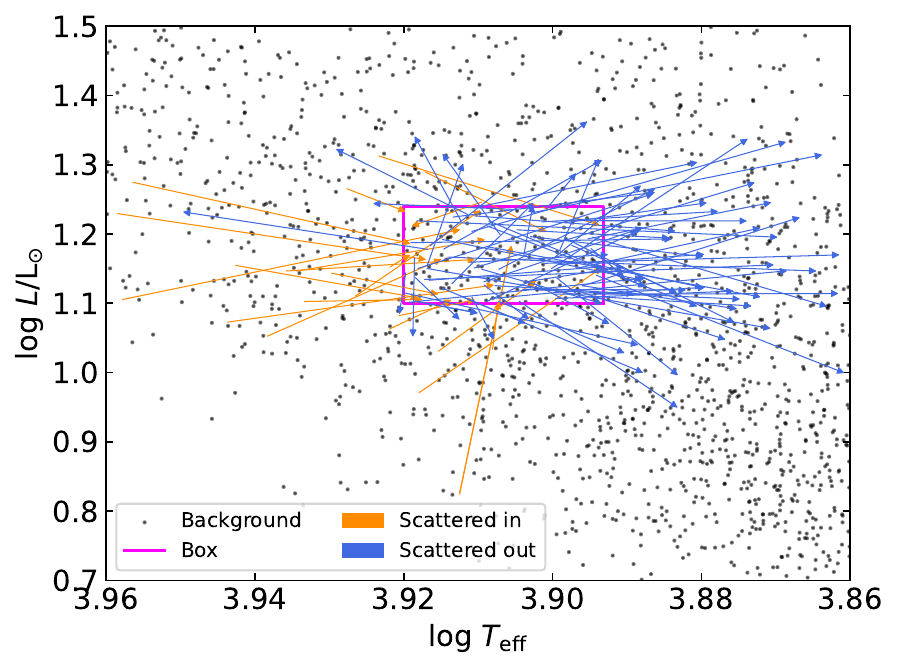}
\caption{The great mixer: various physical effects shift synthesised stars into and out of the observational box (magenta) described in Figures\:\ref{fig:scatter_contributions}\:\&\:\ref{fig:HRD}.
The small grey points show the background distribution of synthesised stars, thinned by a factor 250. Blue arrows show the initial and final positions, corresponding to {\it original} and {\it observed} parameters, for stars that were initially inside the box but got scattered out. In contrast, orange arrows show the stars that were originally outside the box but got scattered in.}
\label{fig:box_scatter}
\end{center}
\end{figure}

The effects of centrifugal deformation are to move stars to the right (and slightly down) in the HR diagram, and the effects of unresolved binaries are to move stars up and right in the HR diagram. These individual contributions were shown in Figs\,\ref{fig:scatter_contributions}\,\&\,\ref{fig:teff_and_logl}. Hence, any population-level analysis that assumes stars are single and non-rotating will be strongly biased. Random observational error further broadens the distribution of stellar properties in any given region in the HR diagram. To indicate the net effect, we consider the impact upon an `observational box' of stars, as shown in Fig.\,\ref{fig:box_scatter}. These processes explain why the binary main sequence that exists for late-type stars \citep{gaiacollaboration2018b} is not evident for early-type stars.

\begin{figure}
\begin{center}
\includegraphics[width=0.48\textwidth]{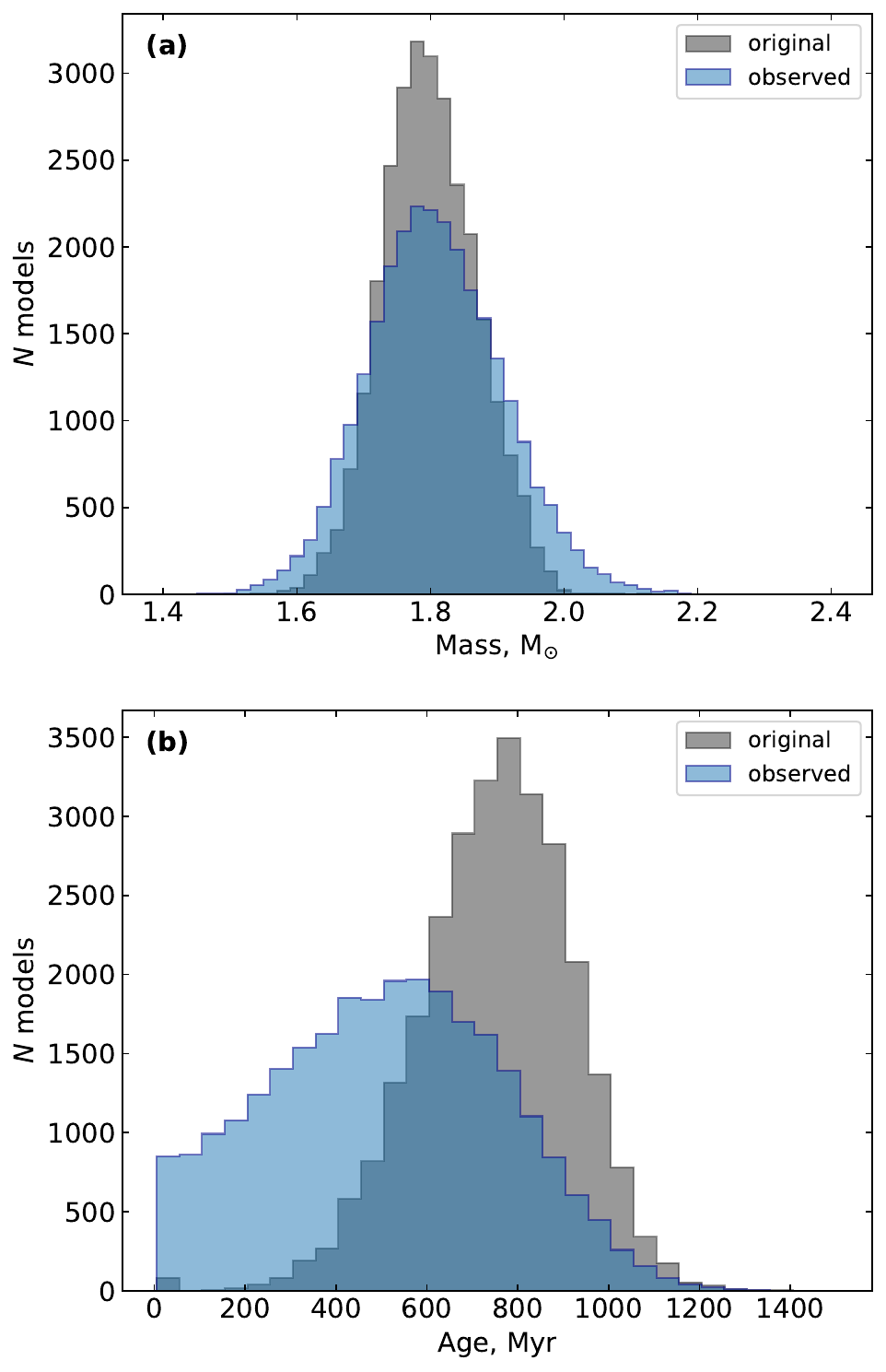}
\caption{The distribution of stellar masses {\bf (a)} and ages {\bf (b)} of stars within an observational box, before (grey) and after (blue) taking into account the effects of binarity, rotation, and random observational uncertainty. The leftmost bin of the {\it original} age distribution consists of pre-MS stars crossing the observational box.}
\label{fig:mass_age_box_scatter}
\end{center}
\end{figure}

It is clear from Fig.\,\ref{fig:box_scatter} that selecting stars from a narrow range of observed $T_{\rm eff}$ and $\log L$ does not succeed in constraining their physical properties to the same extent. The resulting mass and age distributions change markedly (Fig.\,\ref{fig:mass_age_box_scatter}). The mass distribution is 42\% broader, and while the change in its mean of 0.0091\,M$_{\odot}$ is small, the large sample sizes of stars originally ($n=27\,936$) and ultimately ($n=27\,395$) in the observational box makes it statistically significant: Welch's $t$-test, which is like Student's $t$-test but is calculated without the assumption of equal variance of the two samples, gives a $p$\:value of $10^{-35}$ for the null hypothesis of identical means. The age distribution shows more pronounced change. It becomes broader, skewed, and shifted. The central 95\% of the distribution has changed from 406--1048\,Myr to 44--991\,Myr, and its median changed from 759 to 515\,Myr. The magnitude of the effect depends on the $T_{\rm eff}$ and distance from the ZAMS of the observational box, as we shall show in the next subsection. Any inference based on HR diagram position may be biased by the above effects. An obvious example, which we discuss below, is isochrone fitting. Asteroseismic examples, including the use of observations to calibrate theoretical instability strip boundaries and period--luminosity relations, will be the subject of future work.

\subsection{Expected mass and age uncertainties across the A-star main sequence}
\label{ssec:bias_and_uncertainty}

To analyse the dependence of the age bias on position in the HR diagram, we calculated the difference between the {\it observed} and the {\it original} ages, as a fraction of the {\it original} age. Non-rotating isochrones will estimate ages similar to the {\it original} age, i.e., without the effects of rotation and binarity. We use the {\it observed} properties (rather than the {\it binary} properties) because we are making a theoretical comparison between two sets of models, and we want to know the magnitude of the bias and random uncertainty that one can expect from isochrone fitting of a real star. Since the latter would have its own positional uncertainty, we must simulate that uncertainty here by adding the measurement uncertainty (scatter) to the synthetic stars. 

\begin{figure*}
\begin{center}
\includegraphics[width=0.48\textwidth]{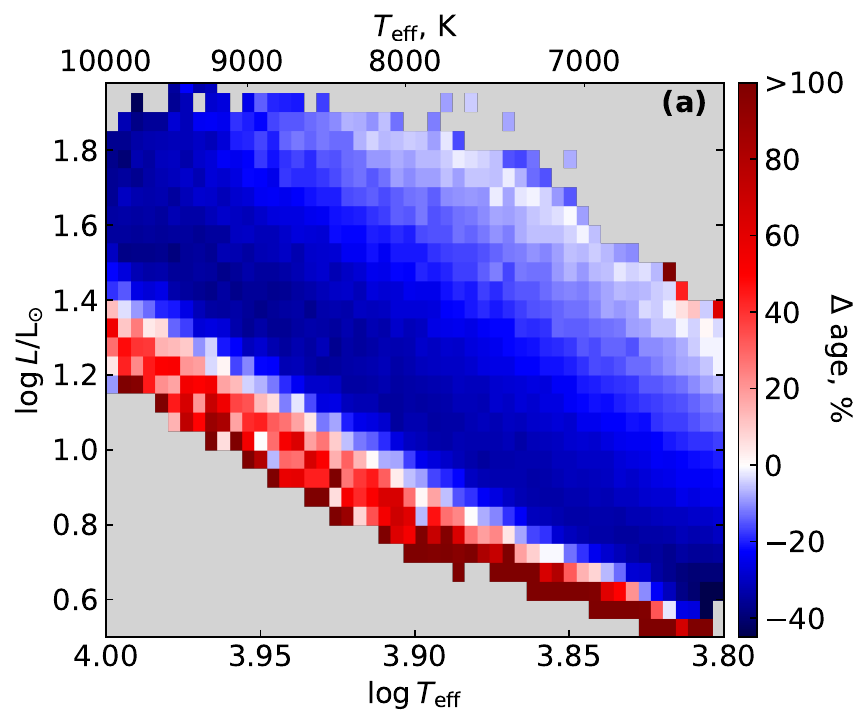}
\includegraphics[width=0.48\textwidth]{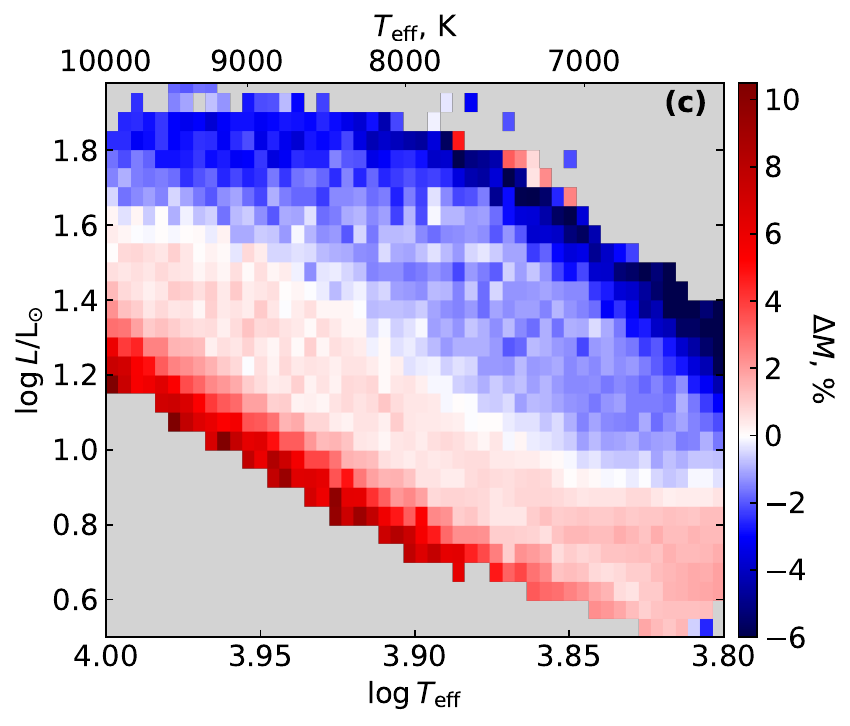}\\
\includegraphics[width=0.48\textwidth]{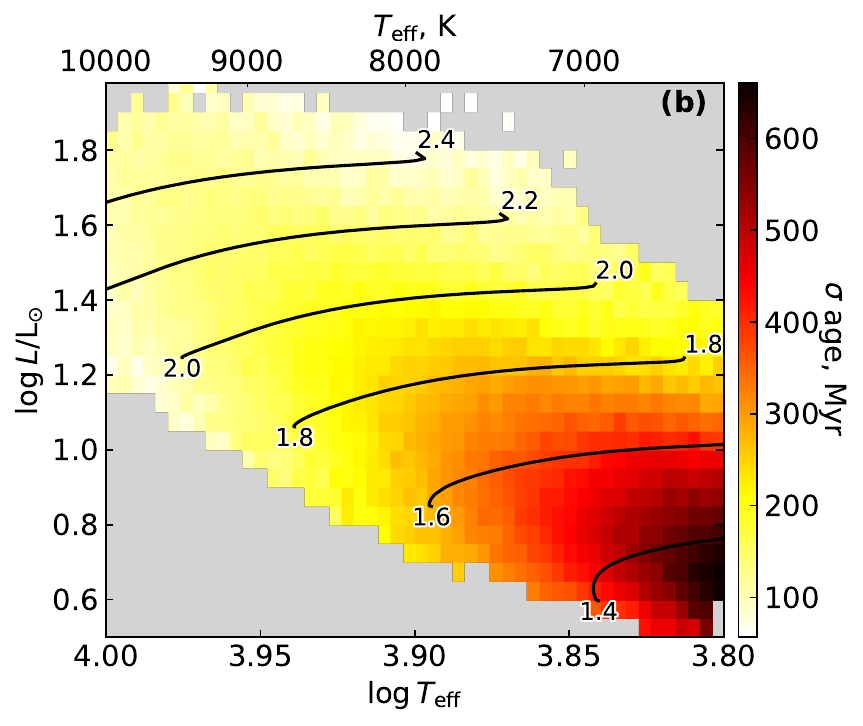}
\includegraphics[width=0.48\textwidth]{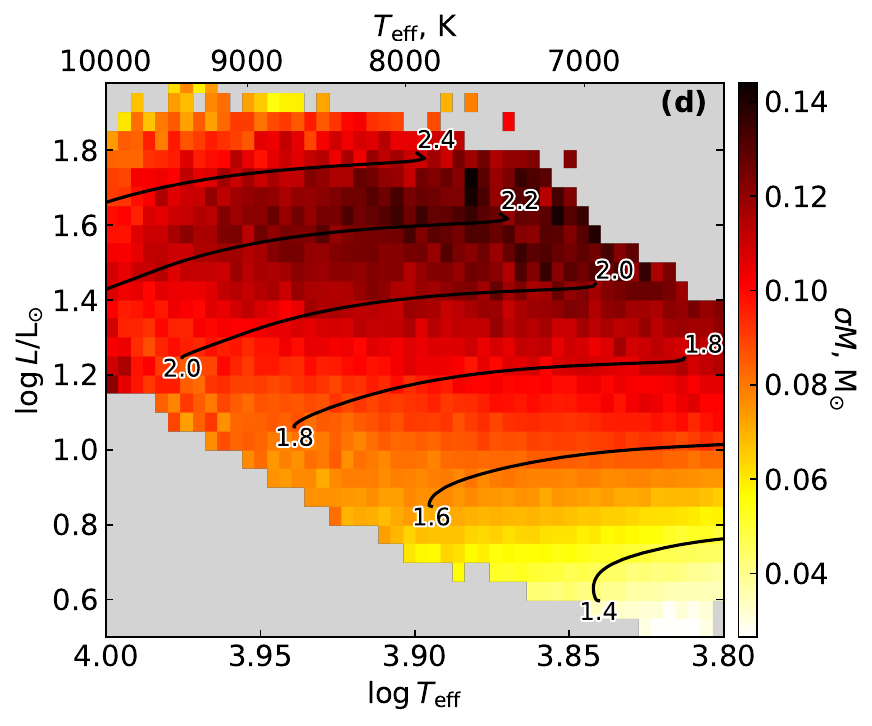}\\
\caption{Ages (left) and masses (right) based on position in the HR diagram are susceptible to bias due to rotation and binarity. This figure shows the typical bias (top) and uncertainty (bottom) one can expect in each parameter from position-based inference such as isochrone fitting. {\bf (a)} Age bias, calculated as the {\it observed} minus {\it original} age, as a percentage of the {\it original} age, for all cells containing 5 or more stars. The colour bar shows the change in mean age in the cell. In red regions, non-rotating isochrones will underestimate stellar ages; in blue regions they overestimate ages. {\bf (b)} The standard deviation of {\it observed} ages in each cell, interpretable as methodological random uncertainty. Solid black lines are evolutionary tracks labelled with their masses in M$_{\odot}$, stretching from the ZAMS to the TAMS, computed with zero rotation and with solar metallicity, for reference. {\bf (c)} Masses are also biased at the ZAMS, where non-rotating isochrones will underestimate them. Although there is a bias at the TAMS, it is smaller and there are other effects to consider (see the text). There is little mass bias across most of the main sequence. {\bf (d)} Methodological random uncertainty on mass, which is greater for more massive stars and rapid rotators.}
\label{fig:age_bias}
\end{center}
\end{figure*}

We used a grid size of 0.004 in $\log T_{\rm eff}$ and 0.05 in $\log L/{\rm L}_{\odot}$, and discarded results from cells with fewer than 5 simulated stars in either the {\it original} or the {\it observed} parameters. Fig.\,\ref{fig:age_bias}a shows that the {\it observed} ages of stars near the ZAMS are much higher than single non-rotating models would predict.\footnote{Although the colour bar is capped at 100\%, 35 of the 167 cells with positive $\Delta$\,age had values exceeding 100\%, and 19 of those exceeded 200\%.} In other words, isochrones will systematically underestimate ages at the ZAMS. This is largely because random measurement uncertainty can scatter older stars into this region, but there are no younger stars around to scatter in to compensate. Meanwhile, the systematic movement of stars upward and to the right also carries away some of the younger population. Away from the ZAMS, ages are systematically overestimated by tens of percent. This is because, for any given cell, younger stars from the lower left will have moved in, while stars originally in the cell will tend to move up and right.

In addition to the bias, there is a methodological random uncertainty, which we calculated as the standard deviation of stellar ages within each cell (Fig.\,\ref{fig:age_bias}b). This random uncertainty is larger for lower mass stars, which have longer lifetimes, but even for higher mass stars the error is seldom less than 100\,Myr. Across the 1022 cells with more than 5 stars, only 103 (10.0\%) of them have random uncertainty $<$100\,Myr. The median is 184\,Myr.

Mass estimates are also biased when similarly inferred from positions in the HR diagram alone (Fig.\,\ref{fig:age_bias}c). Strong biases are seen at the ZAMS, but unlike for ages, substantial biases are not seen in the middle of the main sequence. This is probably because the movement vectors of stars in the HR diagram due to rotation and binarity are almost parallel to evolutionary tracks. Biases also exist at the TAMS, but other effects are important there, and our analysis is not immune to all of them. For instance, we did not include the evolutionary effects of rotation, nor how model parameters such as core overshooting affect the TAMS location \citep[e.g.][]{claret&torres2016}. Conversely, an effect we have accounted for is model sampling density: because evolution is much faster in the post-MS contraction phase, isochrone analyses must account for this difference in prior probability \citep{pont&eyer2005,jorgensen&lindegren2005,dotter2016}. In our synthetic population, this effect is `built in' because models are sampled uniformly -- fewer models will be found there per Myr.

The methodological random uncertainty in mass increases towards higher luminosity (Fig.\,\ref{fig:age_bias}d), except at the top left of the figure where the edge of our grid is apparent (there are no higher-mass stars to scatter down). We attribute this trend to two effects: (i) a fractional dependence on mass, meaning that a given fractional mass uncertainty equates to a higher absolute uncertainty for more massive stars; and (ii) more massive stars rotate more rapidly, on average, hence are located farther from their non-rotating counterparts on the HR diagram than is typically the case for low-mass stars.

\begin{figure*}
\begin{center}
\includegraphics[width=0.98\textwidth]{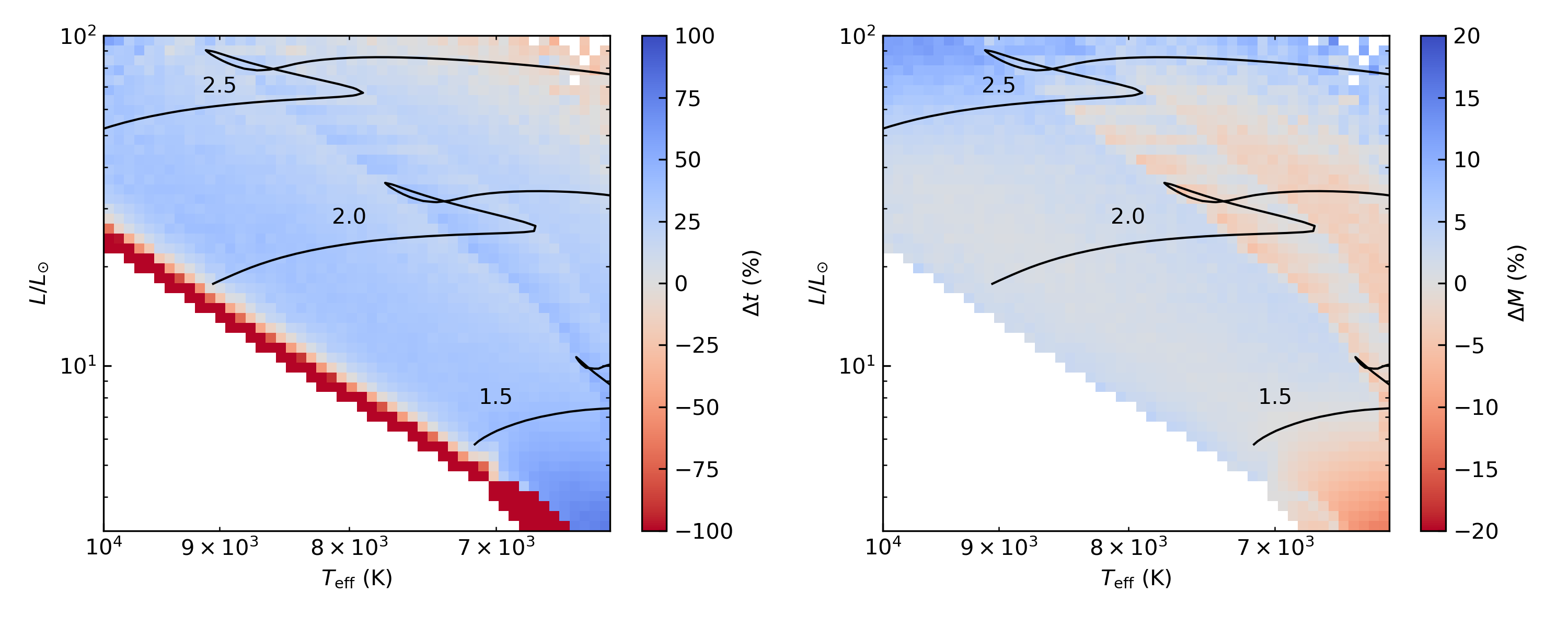}
\caption{Difference in the model stellar properties and those inferred from DSEP isochrones, demonstrating the bias induced by neglecting binarity and rotation for A stars. The colour bars are capped at $\pm$100\%,
and their labels appear upside down compared to Fig.\,\ref{fig:age_bias} because they do not measure the same thing. Fig.\,\ref{fig:age_bias} shows the difference that incorporating rotation, binarity, and measurement error makes. Fig.\,\ref{fig:dsep_isochrones} shows the direct impact on isochrone ages and masses, in particular, that isochrones underestimate ages at the ZAMS and overestimate ages otherwise. After 10\% of the MS lifetime, we determine the bias and scatter to be 31\% and 27\%, respectively.
}
\label{fig:dsep_isochrones}
\end{center}
\end{figure*}

\subsection{Confirmation via fitting of public isochrones}
\label{ssec:isochrone_fitting}
To confirm the above effects, we fitted publicly available isochrones to our simulated stars and evaluated the difference between their physical properties and those returned by the isochrones. We used both DSEP \citep{dotteretal2008} and MIST \citep{dotter2016} isochrones, and importantly, we analysed available isochrones built from models of different rotation rates: 0\% and 40\% of critical\footnote{At source and in the literature citing it, `critical' is used to describe the rotation of these isochrones, but they were calculated using $\omega_{\rm k}$, not $\omega_{\rm c}$.} (`non-rotating' and `rotating' isochrones, hereafter).

Isochrones were matched to the simulated {\it observed} $T_{\rm eff}$ and $\log L$ values by least-squares using the \texttt{kiauhoku} model grid interpolator introduced by \citet{claytoretal2020}. In Fig.\,\ref{fig:dsep_isochrones}, we show the difference between the simulated (known) and the isochrone (inferred) ages and masses. The results are qualitatively similar to those in Fig.\,\ref{fig:age_bias}, with quantitative differences arising from the difference in input physics between the DSEP isochrones and the grid of models upon which our population synthesis is based \citep{gautametal2026}. Moreover, it is imperative to note that our simulation implemented a distribution of metallicities, whereas isochrone analyses typically employ a fixed metallicity, as is replicated here.

Comparisons against non-rotating and rotating MIST isochrones (Figs\,\ref{fig:mist_isochrones}\,\&\,\ref{fig:vcrit_isochrones}), are also qualitatively similar. The latter comparison is important. It shows that using rotating models of a single rotation rate does not remove the bias in the inferred age and mass. This is not because $\omega_{\rm k}=0.4$ is a poor choice for rotating isochrones; on the contrary, this is where the mode of the rotation velocity distribution lies (as we showed in Fig.\,\ref{fig:rotational_distribution}). Instead, it is probably because centrifugal effects scale as $\Omega^2$, so rapid rotators dominate the observed effects. 

We calculate that a 30\% age overestimation for stars in the middle of the main-sequence still remains when using rotating isochrones, and underestimations reaching $\sim$100\% remain near the ZAMS. These biases are accompanied by methodological random uncertainty, as we showed in Sec.\,\ref{ssec:bias_and_uncertainty}, and these effects are not currently accounted for in isochrone fitting in the literature. 

To obtain a summary estimate of the uncertainty of ages inferred with isochrones, we average the age offset across the A-star HR diagram for stars that have surpassed 10\% of their main sequence lifetimes (avoiding small denominator effects). After 10\% of the main sequence lifetime, we find that isochronal ages exhibit a bias of 31\% (isochrone inferred minus simulated) and a scatter of 27\%. These represent averages across the three isochrone grids we tested, but the results are consistent to within 2\% among the grids.

\subsection{Towards realistic parameters and uncertainties with simulation-based inference: application to a hot exoplanet host}
\label{ssec:hd250208}

All is not lost. Stellar parameters can be inferred on a statistical basis by comparing a target star's position on the HR diagram to a reference sample constructed from realistic distributions of stellar properties. In essence, one asks ``what are the properties of nearby stars on the HR diagram?'', and assumes the target star has similar properties.

Here we give a worked example for a hot exoplanet host, TIC\,97568467 = TOI\,2497 = HD\,250208.  This target was chosen without prejudice: it was the first result on the NASA Exoplanet Archive for a ``Published Confirmed'' planet discovered by TESS with stellar $T_{\rm eff}>7000$\,K, at the time of experimental design (2025 Apr). It also helps that the authors \citep{rodriguezetal2023} did a thorough host star characterisation, underpinned by isochrones. They used EXOFASTv2, which uses MIST isochrones, to infer a system age of $1.00^{+0.22}_{-0.19}$\,Gyr, based on measurements of $T_{\rm eff} = 7350^{+270}_{-250}$\,K and $L = 14.7^{+2.0}_{-1.7}$\,L$_{\odot}$ (their table 6). They also inferred a mass of $1.859\pm0.085$\,M$_{\odot}$.

Our Fig.\,\ref{fig:age_bias}a suggests that the age of HD\,250208 is likely to be overestimated by 20\% if non-rotating isochrones were used. It also suggests a methodological random error of $\sim$300 Myr. In the following, we revise the stellar mass and age by comparison with our simulated population. Importantly, for this reference population we use the {\it binary} parameters, i.e. without scatter added, since the target already has $T_{\rm eff}$ and $L$ uncertainties. 

For the purpose of this example, we treat the stellar rotation as unknown, but in Sec.\,\ref{ssec:applet} we show how such information can be included. With Gaia DR4, an observed $v \sin i$ will be available for many targets like this one, and this can inform the outcome. However, note that one does not typically know either $v$ or $i$ independently, so a low $v \sin i$ does not necessarily have a strong bearing on the results. A large $v \sin i$, however, will always imply rapid rotation and a near-equatorial viewpoint, and that the star is more massive and younger than isochrone fitting implies.

\begin{figure}
\begin{center}
\includegraphics[width=0.48\textwidth]{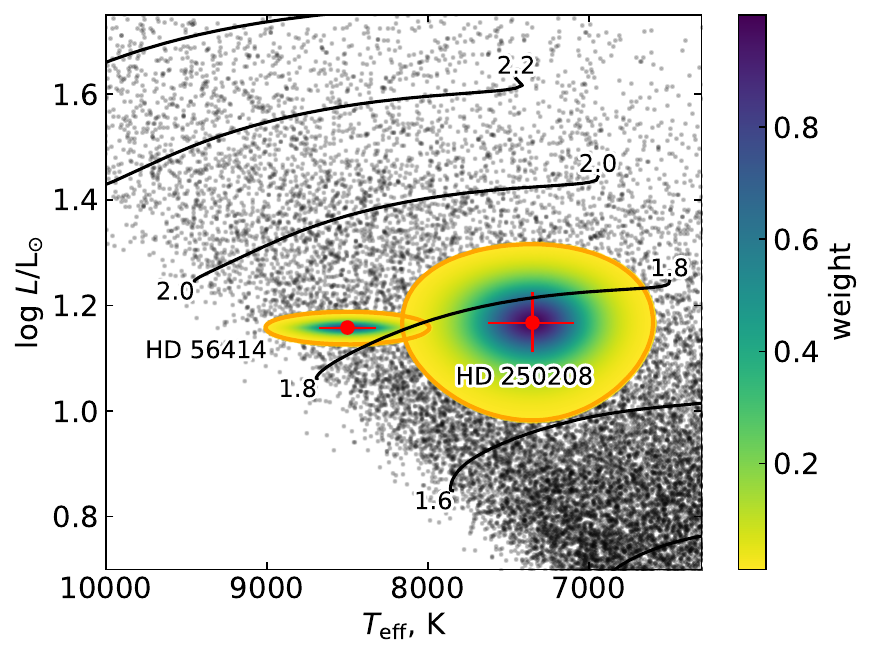}
\caption{The position of the exoplanet hosts HD\,250208 and HD\,56414 are shown with their 1$\sigma$ uncertainties as the red crosshairs. The {\it binary} properties of simulated stars are shown as grey circles, and are thinned to one in every 25 stars. These form the reference sample. Simulated stars with a Mahalanobis distance < 3 from each planet host are shown as coloured points with no thinning; the colour bar is the weight, $w=\exp(-\frac{1}{2}d^2)$. Solid black lines are the evolutionary tracks from Fig.\,\ref{fig:age_bias}.}
\label{fig:mahalanobis_hrd}
\end{center}
\end{figure}

In general, Gaussian uncertainties on a star's position in the HR diagram do not produce Gaussian uncertainties in the underlying stellar parameters, not just because evolutionary tracks lie at an angle to the $T_{\rm eff}$--$L$ axes, but for various evolutionary effects such as more massive stars evolving more quickly and there being fewer of them. Our first step was therefore to use error ellipses, calculating distance on the HR diagram as $$d = \sqrt{\left(\frac{\Delta T_{\rm eff}}{\sigma T_{\rm eff}}\right)^2 + \left(\frac{\Delta L}{\sigma L}\right)^2}$$ also known as the Mahalanobis distance. In this calculation, we retained the asymmetric uncertainties on $T_{\rm eff}$ and $\log L$. These distances were then used as weights, $w = \exp(-\frac{1}{2}d^2)$, for a Gaussian kernel density estimate (KDE) in the desired variable.\footnote{One could implement weights as $w=1/d^2$, but it becomes necessary to cap the maximum weight so that points with a very small Mahalanobis distance do not dominate the KDE. A cap at $d=0.5$ (maximum weight = 4) and normalisation constant of 1/4 gives qualitatively similar results to our formula, and is the alternate form implemented in RAPID.} We considered all simulated points with $d\leq3$ when calculating stellar properties (Fig.\,\ref{fig:mahalanobis_hrd}). For HD\,250208 this selects $n=165\,137$ simulated points for comparison; for HD\,56414 (Sec.\,\ref{ssec:hd56414}), $n=17\,401$.

It is important to refrain from `correcting' for model density. From Fig.\,\ref{fig:mahalanobis_hrd} it is apparent that the model density is sparser to the upper left. This is because we sampled stellar masses from the IMF. Similarly, the absence of stars in some parts of the HR diagram, such as below the ZAMS, is real: there are no models there because there should be no stars there. The model density is the prior, and one should not `correct' for it. Some combinations of stellar parameters are intrinsically less likely, and some are impossible without complex binary evolution.

\begin{figure}
\begin{center}
\includegraphics[width=0.48\textwidth]{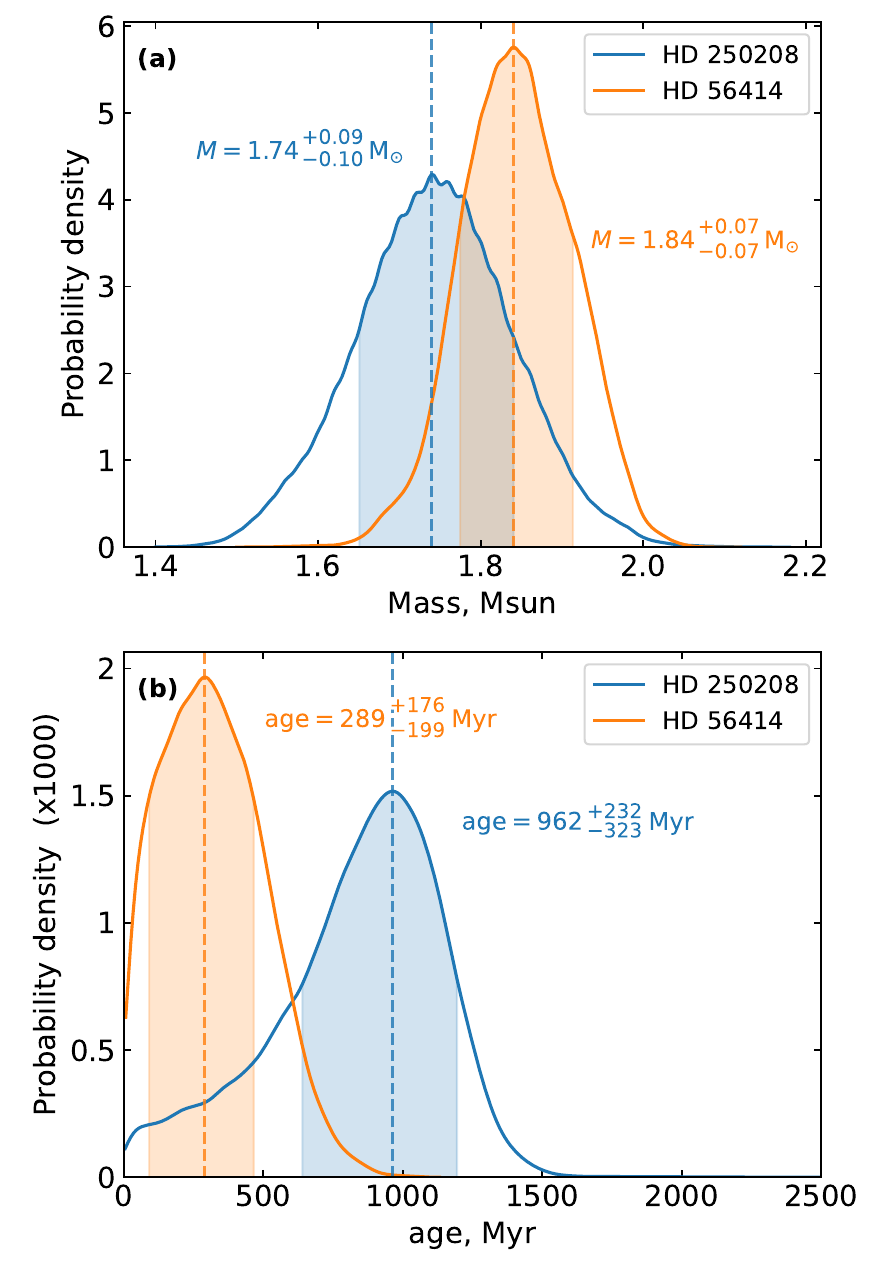}
\caption{The probability distribution function (PDF) for the mass {\bf (a)} and age {\bf (b)} of the exoplanet hosts HD\,250208 and HD\,56414, calculated via kernel density estimation using the Mahalanobis distance as weights. For the mass distributions, the reported values are medians (also shown as dashed vertical lines), and the 1$\sigma$ uncertainties (indicated by the shaded areas). For the age distributions, the mode and highest probability density are used instead for calculating the credible intervals.}
\label{fig:mahalanobis_mass}
\end{center}
\end{figure}

Returning to the specific case of HD\,250208, we first applied our simulation-based inference to the host star mass. The probability density function (PDF) of the KDE is shown in Fig.\,\ref{fig:mahalanobis_mass}a. The median and $\pm1\sigma$ uncertainties were calculated by integrating the PDF, thereby calculating the cumulative distribution function, and evaluating it at percentiles of 50, 84, and 16, respectively. Our calculated mass, $1.74^{+0.09}_{-0.10}$\,M$_{\odot}$, is lower than the published mass by 0.119\,M$_{\odot}$. 

We inferred the stellar age to be $963^{+233}_{-323}$\,Myr, which is consistent with the published age at $1\sigma$, but ours is skewed towards lower ages. Because the age posterior is strongly skewed  (Fig.\,\ref{fig:mahalanobis_mass}b), we report the posterior mode as the central value. The quoted uncertainty corresponds to a highest probability density (HPD) interval that encloses 68.2\% of the posterior probability mass.

Some stellar properties might exhibit strong correlations. The properties of the simulated points in the vicinity of HD\,250208 are essentially posteriors, and it is trivial to make corner plots to see how the system properties could be further constrained by certain observations. We demonstrate this in Sec.\,\ref{ssec:applet}.

\subsection{Application to a younger star}
\label{ssec:hd56414}

Here we provide a second example: HD\,56414. It is a somewhat younger star and was chosen as the hottest exoplanet host on the NASA Exoplanet Archive with a ``Published Confirmed'' planet discovered by TESS. \citet{giacaloneetal2022} give a stellar $T_{\rm eff}$ of $8500\pm150$\,K and a luminosity of $L = 14.39\pm0.34$\,L$_{\odot}$. The luminosity uncertainty is far smaller than that of HD\,250208, producing a very different uncertainties profile (Fig.\,\ref{fig:mahalanobis_hrd}). Since their $T_{\rm eff}$ uncertainty is smaller than 2\%, we used a 2\% uncertainty (170\,K) instead, as per our discussion in Sec.\,\ref{ssec:obs_uncertainty}.

Proceeding with the same approach as for HD\,250208, we infer a stellar host mass for HD\,56414 of $M=1.84^{+0.07}_{-0.07}$\,M$_{\odot}$ and an age of $289^{+176}_{-199}$\,Myr (Fig.\,\ref{fig:mahalanobis_mass}). We quote the posterior median with the central credible interval for mass and the posterior mode with the HPD interval for age. \citet{giacaloneetal2022} quoted an age in their exoplanet discovery paper of $420\pm140$\,Myr. This age came not from isochrone fitting of HD\,56414 directly, but from its membership in Theia\,797 \citep{kounkel&covey2019}, whose age was determined via machine learning and isochrone fitting of the whole association. That age should therefore be rather well determined. It is consistent with ours at $1\sigma$. In cases like this where additional age priors are available, one should not expect that our approach necessarily returns more accurate or more precise ages, but visualising the age posteriors may nonetheless be scientifically useful, up until the target star is over-constrained. That is the topic of Sec.\,\ref{ssec:caveats}.

Note that the mass KDE of HD\,250208 appears slightly featured around the mode. This could reflect that the underlying reference population is ultimately quantised, particularly in mass. This is generally a problem only as the uncertainties on the target star become small, where the model physics and sampling become more important. We discuss this in more detail in Sec.\,\ref{ssec:caveats}. Since the mass distribution for HD\,56414 does not appear similarly featured, quantisation in mass is probably not the correct explanation in this case. Another possible explanation is that the KDE bandwidth is too narrow (we used the default, automatic bandwidth estimation). Users replicating our approach or using RAPID (Sec.\,\ref{ssec:applet}) should take care in choosing an appropriate bandwidth. We allow users to adjust the mass KDE bandwidth with RAPID.

\subsection{Introducing RAPID, the applet}
\label{ssec:applet}

\begin{figure}
    \includegraphics[width=0.97\linewidth]{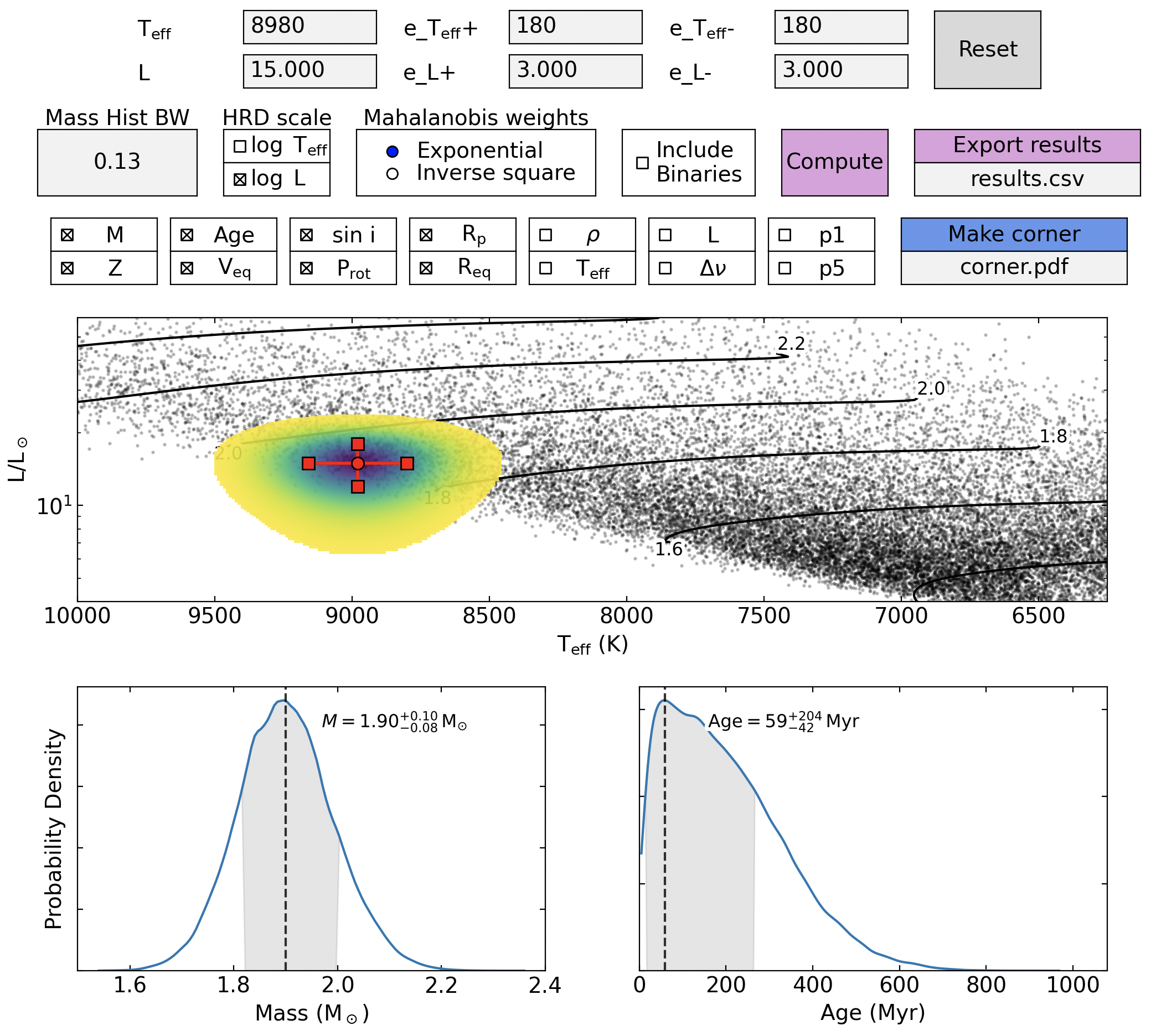}
    \caption{A screen capture from RAPID showing an example target, \mbox{KELT-20}, its corresponding probability density field, and the resulting mass and age posteriors.}
    \label{fig:applet}
\end{figure}

\begin{figure*}
    \begin{center}
    \includegraphics[width=0.98\textwidth]{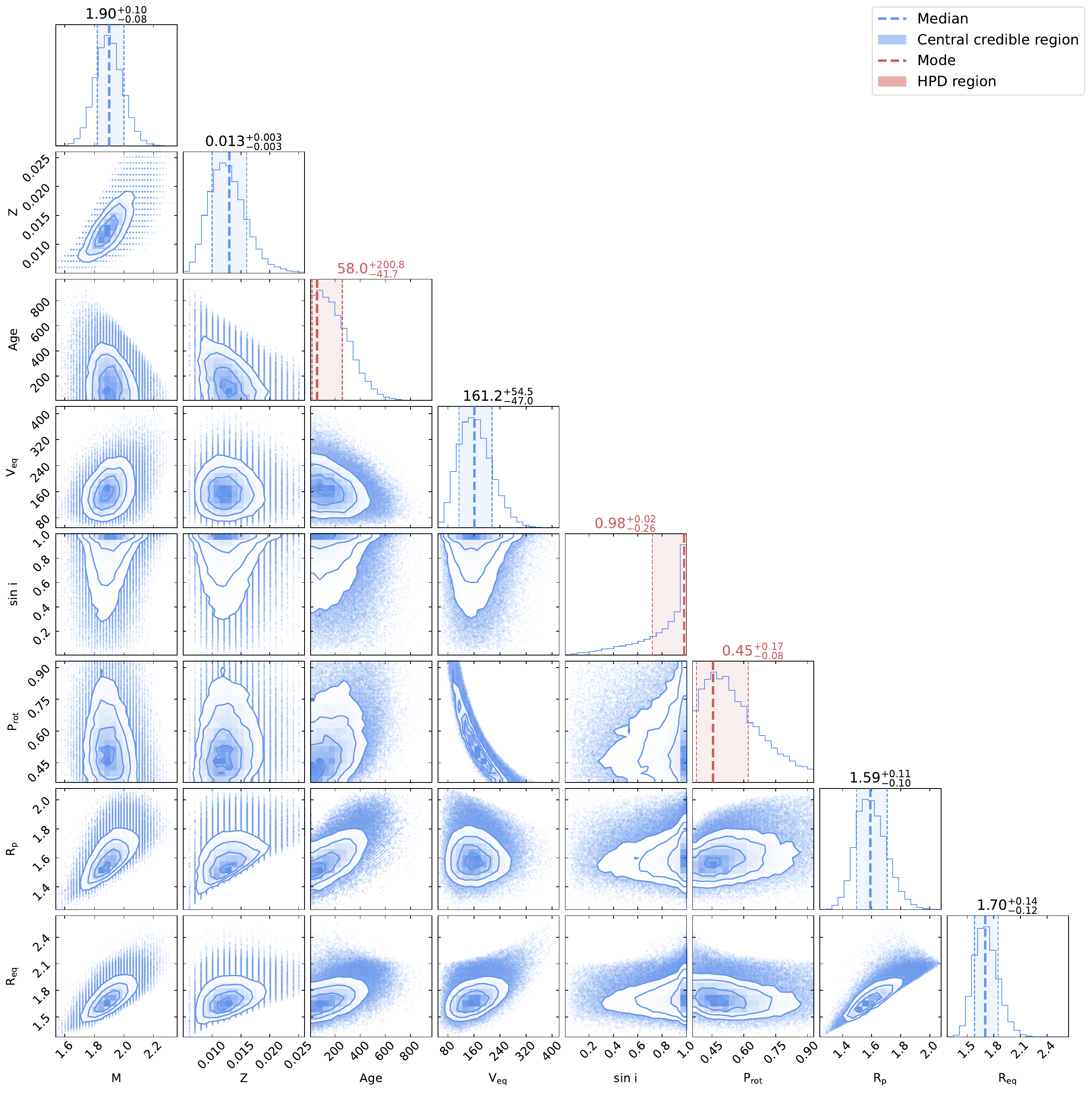}
    \caption{A corner plot of \mbox{KELT-20}, generated using RAPID. The numbers above the diagonal histogram panels represent the median and 1$\sigma$ spread, except for columns where the numbers are displayed in red, which use the mode and Highest Posterior Density (HPD) for the $\pm1\sigma$ credible region, instead. These values are also depicted by the dashed lines on the histogram.}
    \label{fig:corner}
    \end{center}
\end{figure*}

We have created RAPID (Rotation-Aware Probabilistic Inference Dashboard),
an applet that readers can use to quickly infer mass and age distributions of target stars, similarly to our presentation of HD\,250208 and HD\,56414. Figure~\ref{fig:applet} displays a screen-capture from RAPID in use. RAPID implements an interactive visualization and inference tool for exploring stellar properties on an HR Diagram. It allows users to define a target star by specifying its effective temperature and luminosity, along with asymmetric uncertainties in both quantities. The target point and its uncertainty bounds are directly manipulable via draggable graphical elements or editable text fields, enabling intuitive exploration of parameter space. The background displays a population of synthetic stars, which can be toggled between the {\tt full} (default) or {\tt no\_binaries} samples, along with stellar evolutionary tracks that provide astrophysical context for the selected position.

A key feature of RAPID is its use of distance-based weighting to identify stars in the dataset that are consistent with the user-defined target. A generalized Mahalanobis-like metric is computed using the asymmetric uncertainties, and selectable weighting schemes (exponential or inverse-square) are applied to assign likelihood weights to nearby stars. These weights are visualized as a continuous density field around the target out to 3$\sigma$ and are subsequently used to derive posterior distributions for stellar parameters such as mass and age. Gaussian KDEs are employed to produce smooth probability density functions, from which summary statistics are calculated and displayed on ``Compute". We show the posterior median and a central credible interval ($16^{\rm{th}}$ to $84^{\rm{th}}$ percentiles) for mass, and the posterior mode and a HPD region (68.2\% of the probability mass) for age.

In addition to one-dimensional summaries, RAPID supports higher-dimensional exploration through the generation of corner plots for selected stellar parameters. Users can dynamically choose which parameters to include, and the resulting weighted distributions are visualised with annotated summary statistics for each parameter. For parameters with strongly skewed posteriors (e.g.\ age, and when included, $\sin i$ and rotation period), the corner plot titles quote the posterior mode and a HPD interval, highlighted in red. Other parameters are summarised with the median and an equal-tailed central credible interval, shown in blue. Users can also export the 3$\sigma$ results to compute their preferred confidence intervals. Together, these features make RAPID a flexible tool for interactive stellar parameter inference, combining visualisation, uncertainty propagation, and statistical estimation within a single interface. The 3$\sigma$ data exports will download a csv file containing {\it all} columns, as well as $T_{\rm eff}$, $L$, and weights, for further investigation. This can be particularly useful for exploring cuts based on an observed $v \sin i$, [Fe/H], or $\Delta\nu$. 

An example corner plot is shown for KELT-20 (MASCARA-2; HD\,185603) in Fig.\,\ref{fig:corner} ($n=58\,353$), where we took observed properties from \citet{talensetal2018} after inflating the $T_{\rm eff}$ uncertainties to 2\% (Sec.\,\ref{ssec:obs_uncertainty}), and made these RAPID's defaults. A few aspects of the plot are noteworthy. Due to rotation, the posterior distribution of equatorial radius is consistently larger than that of the polar radius, which can be expected to influence exoplanet transit properties, such as durations. Nonetheless, this radius is 2$\sigma$ smaller than the published one ($1.89^{+0.06}_{-0.05}$\,R$_{\odot}$; \citealt{talensetal2018}). Also, we note that our inferred age distribution ($58.0^{+200}_{-41.7}$\,Myr) skews younger than the isochrone age ($200^{+100}_{-50}$\,Myr; \citealt{talensetal2018}), and our mode agrees remarkably well with a new dedicated age analysis for this system ($58\pm5$\,Myr; \citealt{distleretal2026}).
Finally, because KELT-20 is so young (lies close to the ZAMS), certain parameter combinations are inherently less likely. For instance, models with metallicity higher than solar are improbable (and at smaller polar radii, impossible) because evolutionary tracks of higher metallicity lie at systematically cooler $T_{\rm eff}$ and higher $L$. In other words, for such tracks KELT-20 would lie below the ZAMS. These aspects demonstrate the utility of RAPID and underlying data.

Users should take care not to push too far to the edges of the grid, since the resulting statistics will be incomplete. It would be particularly unwise to allow the 3$\sigma$ selection to exceed the input mass range (1.4--2.5\,M$_{\odot}$). Additionally, if the provided uncertainties are too small, the mass KDE can become jagged, reflecting the quantised nature of the input grid. The mass histogram bandwidth is therefore adjustable, should it be necessary, but this will affect the resulting uncertainties. The corner plots, which also apply weights but are not KDE-based, do not have the same problem. Further caveats pertaining to the data are discussed next.

\subsection{Caveats}
\label{ssec:caveats}

Occasionally, targets are so astrophysically important that they are tightly constrained via a multitude of observations. This is more common for nearby stars, and especially pronounced for those accessible to interferometry. Such is the case for HR\,8799, which in addition to interferometric constraints \citep{bainesetal2012} even has its mass constrained from the dynamics of its orbiting planets \citep{sepulveda&bowler2022}, and its age constrained by substellar evolution models of those planets \citep{brandtetal2021}.

Even without dynamical constraints, HR\,8799 has such small observational uncertainties, particularly on luminosity, that the uncertainty in its observables no longer dominate the outcomes. Instead, the model physics has strong bearing. The chosen metallicity distribution of the reference models, including the metal mass fraction chosen for the Sun as the reference star, the helium mass fraction, amount of core overshooting, etc., all influence the result. \citet{sepulveda&bowler2022} conducted a boutique analysis for this target to achieve an ultra-precise age (10--23\,Myr), which is simply out of reach of generalised tools built for just $T_{\rm eff}$ and $L$ as inputs.

We therefore conclude with an advisory to use this tool with its intended purpose in mind, and not to expect to match ultra-precise ages obtained via a multitude of additional, precise priors.

Finally, we address the temptation to use the {\tt no\_binaries} sample in order to achieve smaller error bars on stellar parameters. The purpose of uncertainties is not to have them be as small as possible, but rather, to have them be realistic. Hence, where binaries are not excluded by other means, one should use the {\tt full} sample (`include binaries' button, in RAPID).

\section{Conclusions}
\label{sec:conclusions}

We have synthesised a population of intermediate-mass stars with a solar-neighbourhood metallicity prior and a Salpeter-like initial mass function. To this population, we have added binary star companions, applied the centrifugal effects of rotation to the stellar temperature and luminosity, and assigned random observed inclinations based on an isotropic distribution of inclination axes. The corresponding effects of gravity darkening (or brightening) and luminous companions have been incorporated. Most of these systematically affect observed quantities, meaning that they move stars in a preferred direction in the H--R diagram. We have also added random measurement error, which can move stars in any direction.

The impact of the above effects, compared to non-rotating models without uncertainty, is substantial. Rotation alone can change the observed effective temperature by over 1000\,K, compared to a non-rotating model of the same mass, metallicity and age. We presented an example featuring an observational box, comprising a $T_{\rm eff}$--$\log L$ location and all of the stars within $1\sigma$ of it, whose distribution of masses is 42\% broader, and whose age distribution becomes broader, asymmetrical, and systematically younger, when one accounts for rotation, binarity, and random measurement error.

The implications are wide-reaching. Proximity on the HR diagram does not guarantee similar physical properties such as mass or age, even for stars with comparable metallicities. Moreover, these properties cannot be inferred as precisely as often assumed from $T_{\rm eff}$ and $L$ alone. Strong biases, as well as intrinsic scatter of $\sim$180\,Myr and $\sim$0.1\,M$_{\odot}$, exist when deriving physical parameters from these observables. Because isochrone fitting relies on these same inputs, and often assumes a fixed metallicity, its accuracy and precision for hot stars (above the Kraft break) are significantly lower than commonly believed. We have also demonstrated this by fitting isochrones to our synthetic population.  Excluding stars younger than 10\% of their MS lifetime to avoid small-denominator effects, we found that isochrones systematically overestimate the ages of A stars by 31\% and exhibit a scatter of 27\%. At the ZAMS, isochrones can underestimate ages by a factor of 2 or more.
These findings are especially relevant at a time where automated pipelines apply isochrone-based methods to large stellar datasets with little human oversight.

To further demonstrate the above effects, we modelled ``Published Confirmed'' hot exoplanet hosts using our synthetic population, finding age distributions that skew younger than those published. E.g. for HD\,250208, our one-sigma age confidence intervals extend down to 639\,Myr, compared to the published one that reaches only 810\,Myr. Our analysis of KELT-20 also skews younger than its published isochrone age (58 vs. 200\,Myr for the central value), and is in excellent agreement with a new co-moving based age. If the properties of other hot exoplanet hosts systematically differ from their published values, the effect on exoplanet demographics could be substantial.

The synthetic population is available to the community, and we have provided an applet, RAPID, to interface with it.

Future work will include simulating open clusters, investigating the broader population of exoplanet hosts, and an application to asteroseismology, both in modelling the period--luminosity relationship and in refining the intrinsic boundaries of the $\delta$\,Sct instability strip.

\section*{Acknowledgements}

SJM was supported by the Australian Research Council through Future Fellowship FT210100485. We thank Arif Solmaz, Daniel Huber, Daniel Reese, George Zhou, Luke Bouma, and Tomasz Rozanski for discussions, along with the USyd--UniSQ $\delta$\,Sct Research Group.


\section*{Data Availability}

We have made the synthetic population and the RAPID applet publicly available in a \href{https://github.com/SimonJMurphy/RAPID}{GitHub repository}. We provide our two datasets, {\tt full} and {\tt no\_binaries}, in .csv and .feather file formats there.


\bibliographystyle{mnras}
\interlinepenalty=10000
\bibliography{sjm_bibliography} 


\appendix

\section{Calculation of effective temperature for a deformed star}
\label{app:teff}

Rotation deforms the stellar surface through centrifugal acceleration, which in turn affects the relation between luminosity, surface area, and the global effective temperature. We introduced this effect in Sec. \ref{sssec:centrifugal}. Here, we compare different methods for calculating the global effective temperature of a centrifugally distorted star. These include:

\begin{enumerate}
    \item \textbf{Effective temperature from {\sc mesa}} ($T_\mathrm{eff,{MESA}}$): 
    The effective temperature reported by {\sc mesa} is set through the atmospheric boundary conditions. As noted in \citet{paxtonetal2013} and \citet{jermynetal2023}, {\sc mesa} treats rotation in one dimension using the shellular approximation, such that centrifugal distortion enters the stellar structure equations through correction factors. In the rotation formalism described by \citet{paxtonetal2019}, these correction factors are computed using analytical fits to Roche equipotentials.

    In the underlying {\sc mesa} models used in this work, the atmosphere is described using the Eddington $T$--$\tau$ relation,
    \begin{eqnarray}
        T^4(\tau)=\frac{3}{4}T_\mathrm{eff}^4\left(\tau+\frac{2}{3}\right).
    \end{eqnarray}
    With the photosphere boundary taken at $\tau=2/3$, this gives $T(\tau=2/3)=T_\mathrm{eff}$. For this adopted atmosphere prescription, the temperature at the photosphere boundary equals the effective temperature. The corresponding {\sc mesa} value can therefore be expressed as
    \begin{eqnarray}
        T_\mathrm{eff,{MESA}} = \left(\frac{L}{4\pi R_{\rm phot}^2 \sigma_{\mathrm{SB}}}\right)^{1/4},
    \end{eqnarray}
    where $L$ is the stellar luminosity, $\sigma_{\mathrm{SB}}$ is the Stefan--Boltzmann constant, and $R_{\rm phot}$ is the radius of the layer at $\tau=2/3$.

    For our population synthesis models, we use an interpolator based on rotating {\sc mesa} models (see Sec.\,\ref{sssec:centrifugal}) to compute $T_\mathrm{eff,{MESA}}$ at the sampled rotation rates. This serves as the benchmark effective temperature against which we compare the other definitions. 
    
    \item \textbf{Oblate spheroid effective temperature} ($T_\mathrm{eff,oblate}$): Temperature calculated using the surface area of the corresponding oblate spheroid:
    \begin{eqnarray}
    T_\mathrm{eff,oblate} = \left(\frac{L}{S_\mathrm{oblate} \sigma_{\mathrm{SB}}}\right)^{1/4},
    \end{eqnarray}
    where $L$ is the luminosity, $\sigma_{\mathrm{SB}}$ is the Stefan-Boltzmann constant, and $S_\mathrm{oblate}$ is the surface area of the oblate spheroid given by:
    \begin{eqnarray}
        S_\mathrm{oblate} = 2\pi R_\mathrm{eq}^2\left[1 + \frac{1-e^2}{e}\tanh^{-1}(e)\right],
    \end{eqnarray}
        with eccentricity $e = \sqrt{1-(R_\mathrm{p}/R_\mathrm{eq})^2}$.

    \item \textbf{Volumetric-equivalent radius temperature} ($T_\mathrm{eff,volumetric}$): Temperature calculated using the radius of a sphere with the same volume as the oblate star:
    \begin{eqnarray}
        R_\mathrm{vol} &=& (R_\mathrm{eq}^2 R_\mathrm{p})^{1/3},\\
        T_\mathrm{eff,volumetric} &=& \left(\frac{L}{4\pi R_\mathrm{vol}^2 \sigma_{\mathrm{SB}}}\right)^{1/4}.
    \end{eqnarray}

    \item \textbf{Roche-approximation effective temperature} ($T_\mathrm{eff,Roche}$): Temperature calculated using the surface area given by the analytical approximation to the Roche equipotential adopted in the {\sc mesa} rotation formalism:
    \begin{eqnarray}
        T_\mathrm{eff,Roche} &=& \left(\frac{L}{S_\mathrm{Roche}\sigma_{\mathrm{SB}}}\right)^{1/4},
    \end{eqnarray}
    where $L$ is the stellar luminosity, $\sigma_{\mathrm{SB}}$ is the Stefan--Boltzmann constant, and
    \begin{eqnarray}
        S_\mathrm{Roche} = 4\pi R_\mathrm{eq}^2 \left(1 - \frac{\omega_{\rm k}^2}{3} + 0.08525\,\omega_{\rm k}^4 - 0.04908\,\omega_{\rm k}^6 \right), 
    \end{eqnarray}
    with $R_\mathrm{eq}$ the equatorial radius and $\omega_{\rm k} = \Omega/\Omega_{\rm K}$. This expression provides an approximate surface area for a centrifugally distorted star in the Roche model. It comes from \citet{paxtonetal2019}, where the error on $S_\mathrm{Roche}$ is described as being $<$0.01\%.
\end{enumerate}

\begin{figure}
    \centering
    \includegraphics[width=0.995\linewidth]{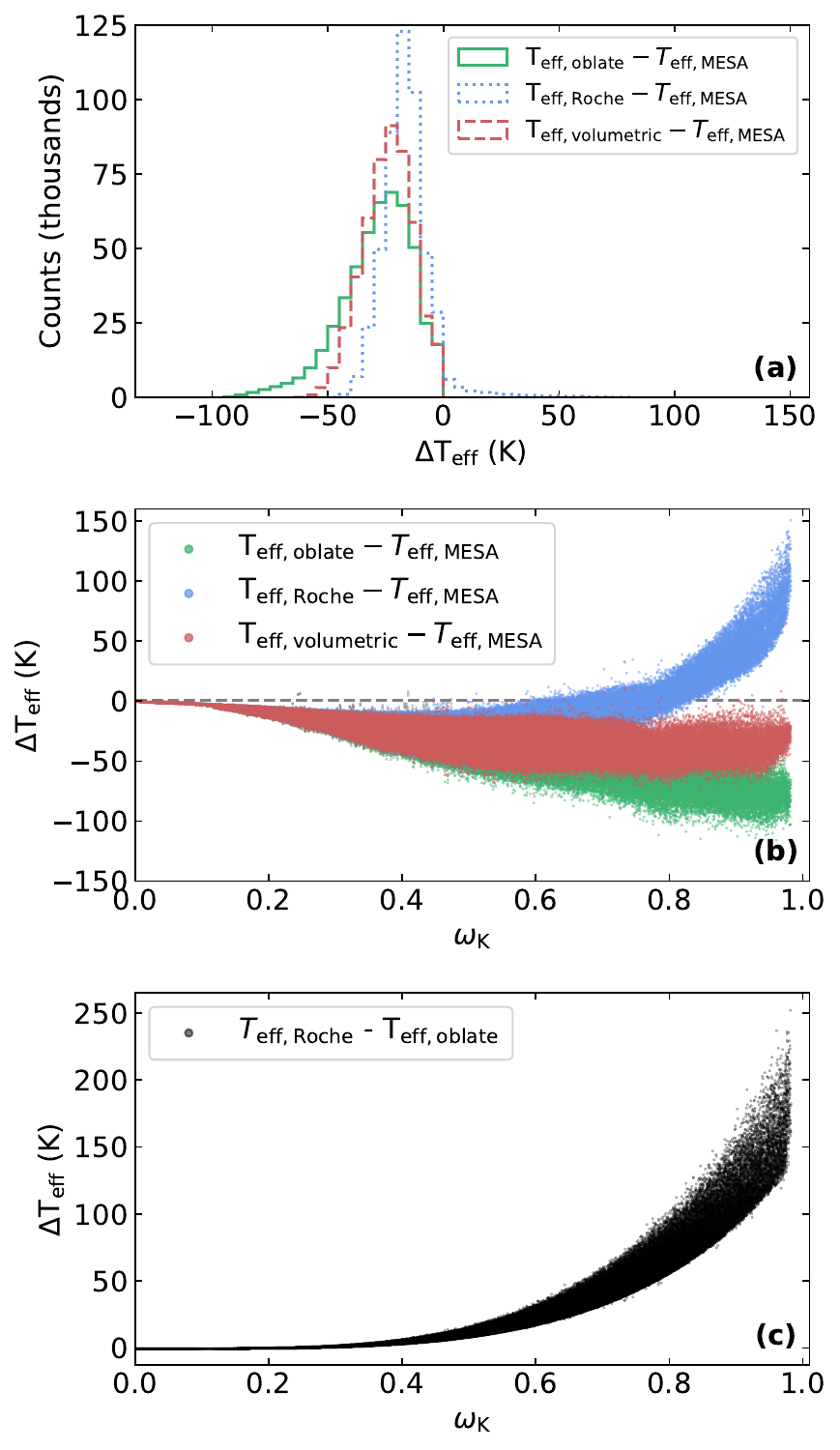}
    \caption{Comparison of effective temperature calculations for rotating stars. {\bf (a)}\,The distribution of temperature differences between various calculation methods. The histograms display the deviation of $T_\mathrm{eff,Roche}$ (blue), $T_\mathrm{eff,volumetric}$ (red), and $T_\mathrm{eff,oblate}$ (green) from the {\sc mesa}-reported temperature $T_\mathrm{eff,MESA}$. {\bf (b)}\,The temperature differences between these methods and {\sc mesa} for our population synthesis models as a function of rotation rate. {\bf (c)}\,The temperature differences between the Roche shape and the oblate spheroid.}
    \label{fig:teff_diffferences}
\end{figure}

For these calculations, we followed a similar methodology as described in Sec. \ref{sssec:omega} to get the critical rotation rate and the equatorial radius of a centrifugally deformed star. The luminosities for (ii), (iii) and (iv) were derived from an interpolation function based on {\sc mesa} rotating models, as in Sec.\,\ref{sssec:centrifugal}.

Figure\:\ref{fig:teff_diffferences} shows the differences between the various effective temperature estimates. Relative to $T_\mathrm{eff,MESA}$, the mean offsets are $T_\mathrm{eff,oblate} - T_\mathrm{eff,MESA} = -28.8$\,K, $T_\mathrm{eff,volumetric} - T_\mathrm{eff,MESA} = -23.9$\,K, and $T_\mathrm{eff,Roche} - T_\mathrm{eff,MESA} = -15.6$\,K. The corresponding standard deviations are 16.02\,K, 10.86\,K, and 12.90\,K. While all methods yield similar effective temperatures for slowly to moderately rotating young stars, the differences grow with increasing $\omega_{\rm k}$. Yet the magnitude of these differences is small compared to the underlying effect: centrifugal distortion can cause $T_{\rm eff}$ reductions of $\sim$1000\,K for the most rapid rotators, whereas the shape based methodological uncertainty we identify here is an order of magnitude smaller. Throughout this work, we adopted the effective temperature calculation for the Roche shape, $T_\mathrm{eff,Roche}$.

\section{Additional isochrone comparisons}
\label{app:isochrones}

\begin{figure*}
\begin{center}
\includegraphics[width=0.98\textwidth]{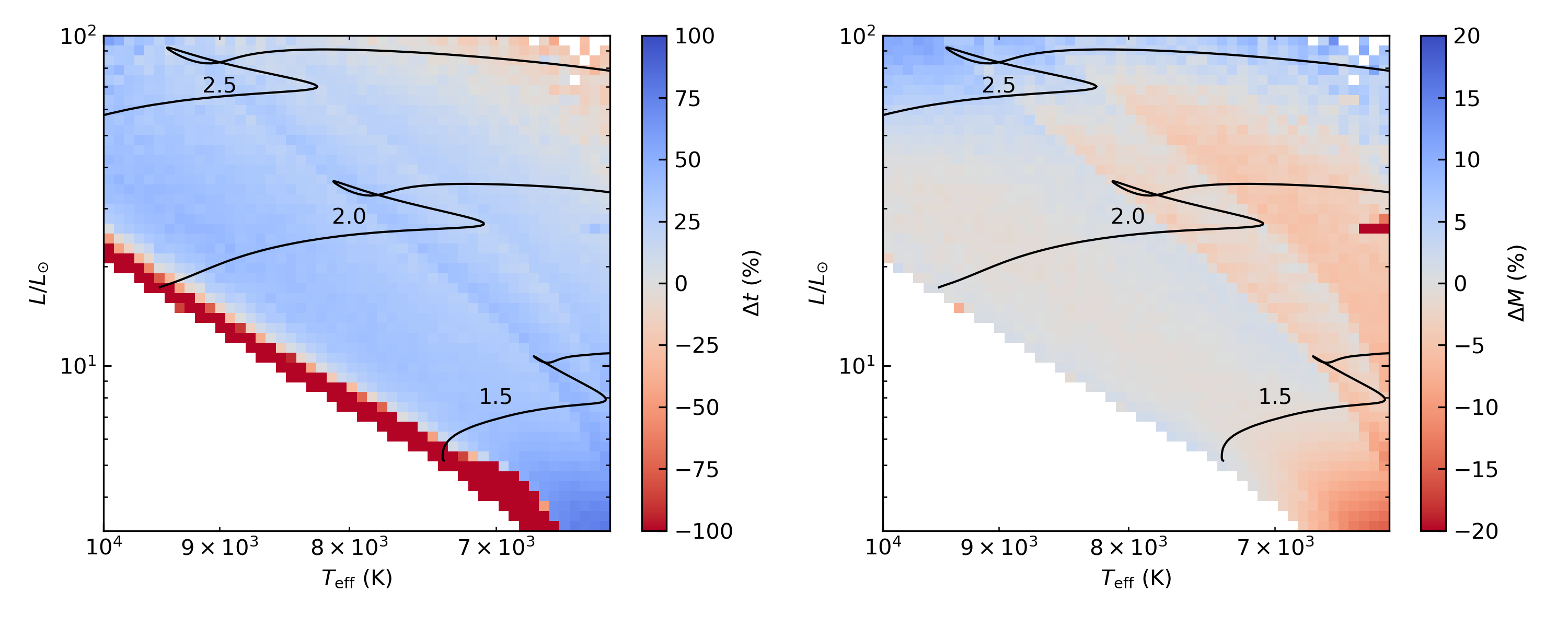}\\
\caption{As Fig.\,\ref{fig:dsep_isochrones}, but for MIST isochrones constructed from non-rotating models.}
\label{fig:mist_isochrones}
\end{center}
\end{figure*}

\begin{figure*}
\begin{center}
\includegraphics[width=0.98\textwidth]{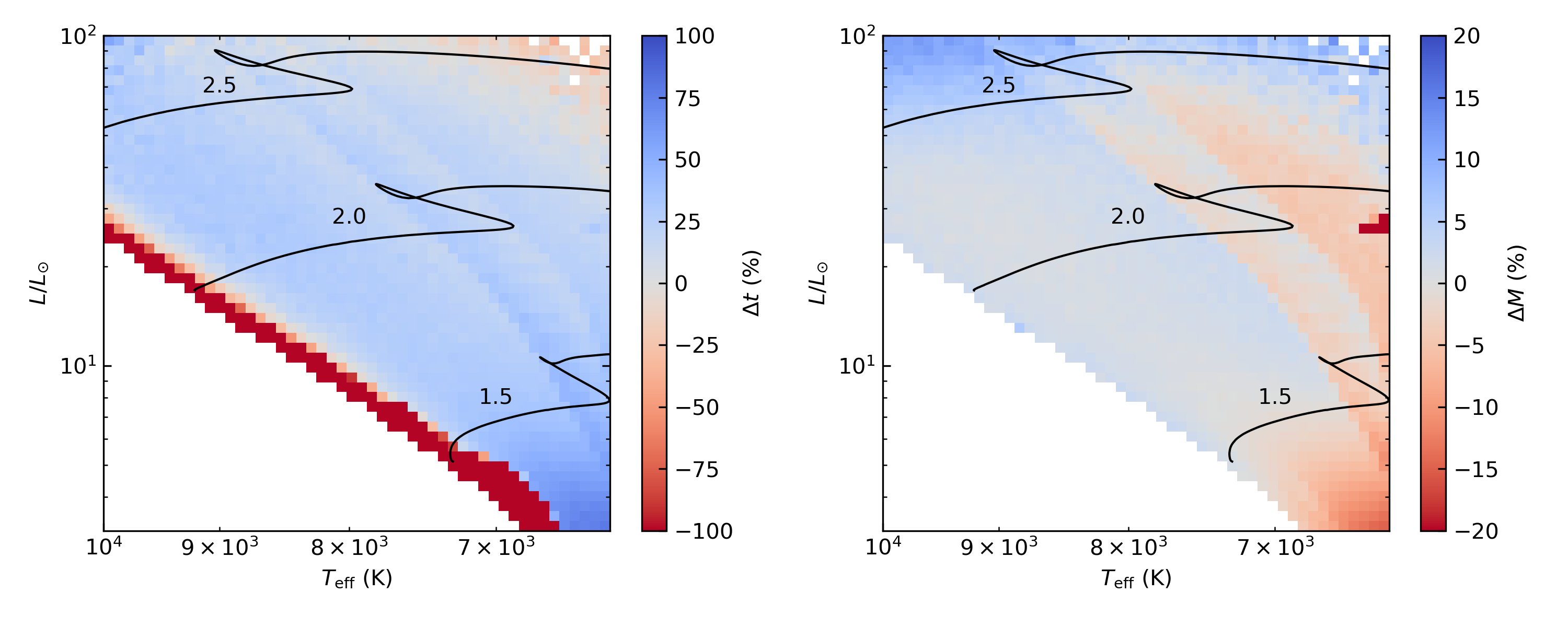}\\
\caption{As Fig.\,\ref{fig:dsep_isochrones}, but for MIST isochrones constructed from models rotating at 40\% of their critical rotation rate. }
\label{fig:vcrit_isochrones}
\end{center}
\end{figure*}

Here we provide additional figures demonstrating that biases in inferred masses and age remain when using different public isochrone datasets, including those constructed from rotating models (Figs\,\ref{fig:mist_isochrones}\,\&\,\ref{fig:vcrit_isochrones}). The differences in these figures are subtle. Their main differences are the location of the ZAMS and the TAMS at the higher-mass end (both are hotter for the non-rotating models), and in the shading of the age differences (larger differences are encountered for non-rotating models). Thus, the use of rotating isochrones only slightly diminishes the bias in inferred ages; it still remains at the 30\% level in the middle of the main-sequence, and remains at $\sim$100\% near the ZAMS.

\bsp	
\label{lastpage}
\end{document}